\documentclass[trackchanges,twocolumn]{aastex701}

\usepackage{amsmath}

\usepackage{siunitx}
\DeclareSIUnit{\fm}{\femto\m}

\newcommand{\sro}[0]{\texttt{SRO}}
\newcommand{\apr}[0]{\texttt{APR}}
\newcommand{\hws}[0]{\texttt{HWS}}
\newcommand{\code}[0]{\texttt{FINEST}}
\newcommand{\rkin}[0]{\text{kin}}
\newcommand{\rpot}[0]{\text{pot}}

\begin{document}

\title{Finite-temperature ab-initio-Informed Nuclear Equation-of-State Tables (\code): \\
Nuclear matter uncertainties in core-collapse supernovae}

\author[orcid=0009-0009-4357-0373]{Finia P. Jost}
\affiliation{Institut für Kernphysik, Technische Universität Darmstadt, Schlossgartenstr. 2, 64289 Darmstadt, Germany}
\email[show]{finia.jost@tu-darmstadt.de, fp.jost@nbi.ku.dk}  

\author[orcid=0000-0003-0849-7691]{Andr\'e da Silva Schneider} 
\affiliation{Departamento de Física, Universidade Federal de Santa Catarina, Florianópolis, SC 88040-900, Brazil}
\email{schneider.andre@ufsc.br}

\author[orcid=0000-0002-6995-3032]{Almudena Arcones}
\affiliation{Institut für Kernphysik, Technische Universität Darmstadt, Schlossgartenstr. 2, 64289 Darmstadt, Germany}
\affiliation{GSI Helmholtzzentrum für Schwerionenforschung GmbH, Planckstr. 1, 64291 Darmstadt, Germany}
\affiliation{Max-Planck-Institut für Kernphysik, Saupfercheckweg 1, 69117 Heidelberg, Germany}
\email{almudena.arcones@physik.tu-darmstadt.de}

\author[orcid=0000-0001-8027-4076]{Achim Schwenk}
\affiliation{Institut für Kernphysik, Technische Universität Darmstadt, Schlossgartenstr. 2, 64289 Darmstadt, Germany}
\affiliation{ExtreMe Matter Institute EMMI, GSI Helmholtzzentrum f\"ur Schwerionenforschung GmbH, 64291 Darmstadt, Germany}
\affiliation{Max-Planck-Institut für Kernphysik, Saupfercheckweg 1, 69117 Heidelberg, Germany}
\email{schwenk@physik.tu-darmstadt.de}

\author[orcid=0000-0002-8228-796X]{Evan O'Connor}
\affiliation{The Oskar Klein Centre, Department of Astronomy, Stockholm University, AlbaNova, SE-106 91 Stockholm, Sweden}
\email{evan.oconnor@astro.su.se}

\author[0000-0002-8984-0211]{Gerard Nav\'o}
\affiliation{Departament d'Astronomia i Astrof{\'i}sica, Universitat de Val{\`e}ncia, Av/ Vicent Andr\'es Estell\'es, 19, 46100 Burjassot, Spain}
\email{gerard.navo@uv.es}

\author[0000-0001-9252-823X]{Yeunhwan Lim}
\affiliation{Department of Physics, Yonsei University, Seoul 03722, South Korea}
\email{ylim@yonsei.ac.kr}

\begin{abstract}

We construct new finite-temperature, composition-dependent equation of state (EOS) tables for astrophysical simulations by implementing a recently developed energy-density functional. The functional is constrained by ab initio nuclear matter calculations and neutron-star observations. A distinguishing feature of this functional is a non-monotonic density dependence of the nucleon effective mass, with an increase beyond saturation density.
Using these EOS tables, we investigate the impact of current uncertainties in nuclear matter properties on core-collapse supernova (CCSN) evolution with spherically symmetric simulations employing an effective treatment of convective mixing. The magnitude of the EOS effects depends on the progenitor compactness. Within current constraints, variations in incompressibility, symmetry energy, and high-density pressure produce only modest differences during the early post-bounce evolution. In contrast, the effective mass dependence has a much stronger influence on proto-neutron star contraction and the explosion dynamics in a stellar model with high core compactness. Its non-monotonic density dependence leads to a qualitatively different contraction history and can alter the explosion outcome compared to mean-field effective-mass behaviors.
Therefore, among the nuclear matter properties considered, the density dependence of the nucleon effective masses has by far the largest impact on CCSN dynamics. This highlights the importance of realistic effective-mass parameterizations in nuclear EOSs used in astrophysical simulations.

\end{abstract}

\section{Introduction}
\label{sec:Introduction}

Astrophysical simulations require a description of the composition and properties of matter under a wide range of conditions. Simulations of core-collapse supernovae (CCSNe) or neutron star mergers (NSMs) that involve neutron stars especially face this challenge. The equation of state (EOS) for the hot and dense nuclear matter in newly born or colliding neutron stars is not yet fully understood. Beyond the uncertainties of the EOS in neutron stars, the challenge of an EOS for astrophysical applications is to describe matter consistently over a wide range of mass density $\rho$, temperature $T$, and electron (or proton) fraction $Y_e$: $\SI{e4}{\gram\per\cubic\centi\metre} \lesssim \rho \lesssim \SI{e15}{\gram\per\cubic\centi\metre}$, $\SI{0.1}{\mega\electronvolt} \lesssim T \lesssim \SI{100}{\mega\electronvolt}$, $0.01 \lesssim Y_e \lesssim 0.6$. 
This extensive range of conditions includes matter in very different phases, from uniform nuclear matter through so-called pasta phases to the formation of nuclei. The EOSs that are commonly used in CCSN and NSM simulations are based on phenomenological models, e.g., employing Skyrme energy-density functionals (EDFs) \citep{lattimer_generalized_1991,schneider_open-source_2017} or relativistic mean-field descriptions \citep{shen_relativistic_1998, shen_new_2011, hempel_new_2012, steiner_core-collapse_2013}. However, many of the EOSs that are typically employed in astrophysical simulations are not consistent with new constraints from observations and/or from ab initio nuclear matter calculations.

There have been many advances in nuclear matter calculations based on chiral effective field theory (EFT) interactions (see, e.g., \citealp{hebeler_equation_2013,Tews:2012fj,Lynn:2015jua,drischler_chiral_2019,carbone_setting_2020,Keller:2022crb,Tews:2024owl,alp_equation_2025,drischler_many-body_2026}), from nuclear experiments (see, e.g., \citealp{wiescher_critical_2012, meisel_dripline_2016, Huth:2021bsp, lattimer_constraints_2023, sorensen_dense_2024, kumar_theoretical_2024}), as well as observations of heavy-mass neutron stars \citep{Demorest:2010bx,antoniadis_massive_2013,cromartie_relativistic_2020,Fonseca21}. In addition, neutron star radius constraints from LIGO/Virgo through gravitational-wave measurements of tidal deformabilities \citep{abbott_gw170817_2017,abbott_properties_2019,Abbott:gw190425} and from NICER using pulse-profile modeling \citep{riley_nicer_2019,miller_psr_2019,Riley21,Miller21,Choudhury:2024xbk,Salmi:2024aum,dittmann_more_2024,Vinciguerra:2023qxq,Miller:2025qfq} have resulted in new EOS constraints (see, e.g., \citealp{Rutherford:2024srk,Koehn:2025,Ng:2025,Somasundaram:2024ykk,Mauviard:2025dmd,Mendes:2026mgc}).

In this work, we use the EOS framework developed by \cite{huth_new_2021}, which takes into account state-of-the-art constraints from theoretical calculations at low and higher densities, as well as observational constraints from heavy pulsar mass measurements, LIGO/Virgo, and NICER. It also includes ab initio results for the density dependence of the nucleon effective masses from \citet{carbone_ab_2019}. We have incorporated these novel parameterizations of the energy density and nucleon effective mass into an updated version of the \sro\ code \citep{schneider_open-source_2017, schneider_akmal-pandharipande-ravenhall_2019}, called \code\ (Finite-temperature ab-initio-Informed Nuclear Equation-of-State Tables). We provide an extensive set of new EOS tables to the astrophysics community, systematically varying nuclear matter properties within their constraints.

These EOS tables are equally applicable in CCSN simulations, NSM simulations, or other astrophysical applications involving neutron stars. In this work, we focus on the impact on CCSNe and propagate the nuclear matter uncertainties to the explosion.

There has been a long history of studies exploring the impact of the EOS on CCSN dynamics and observables (see, e.g., \citealp{burrows_prompt_1985, sumiyoshi_postbounce_2005, marek_equation--state_2009, oconnor_black_2011, hempel_new_2012, couch_dependence_2013, suwa_importance_2013, richers_equation_2017, nagakura_simulations_2018, ghosh_pushing_2022, Boccioli:2021fjh, rusakov_exploration_2026}). These studies demonstrated the importance of the EOS in several aspects of CCSN simulations: The central density and electron fraction at core bounce, the rate and energy at which neutrinos are radiated, and how much energy they feed to the shock. Consequently, also partly the success of explosions, the ejecta mass and composition, and explosive nucleosynthesis yields are affected by the EOS. These studies generally show a trend that simulations with a softer EOS (where the pressure increases less with density) are more prone to yield explosions (see, e.g., \citealp{janka_neutrino-driven_2005, couch_dependence_2013}). Typically, rather different EOS schemes that are popular in the CCSN community have been compared and this is still a common approach (see, e.g., \citealp{couch_dependence_2013, suwa_importance_2013, Boccioli:2021fjh}). Recently, there have been new approaches of systematically varying nuclear matter properties within one theoretical framework. This allows to understand and constrain the isolated effects of a single nuclear property. \cite{yasin_equation_2020} and \cite{schneider_equation_2019} performed this systematic variation using Skyrme EDFs and showed that the parameter with the largest impact on the first second of the explosion, given present uncertainties, are the nucleon effective masses, which govern thermal effects. In relativistic mean-field EOS models, the nucleon effective mass also has a similar impact on CCSN simulations as shown recently \citep{sumiyoshi_influence_2026}. The importance of thermal effects in CCSN dynamics had already been recognized from comparisons of finite-temperature EOS models, where differences in the thermal behavior of the EOS were found to affect proto-neutron star (PNS) contraction, neutrino emission, and shock evolution (see, e.g., \citealp{sumiyoshi_postbounce_2005, hempel_new_2012}). The systematic variations of nuclear matter properties demonstrated that this sensitivity can be traced to the nucleon effective masses, which control the thermal nucleonic contributions to the EOS (\citealp{schneider_equation_2019, yasin_equation_2020}; see also \citealp{eggenberger_andersen_equation--state_2021}). 

We employ a similar approach with the new EDF of \cite{huth_new_2021}, varying nuclear matter properties individually and investigating their impact on the first seconds of the explosion. In our spherically symmetric CCSN simulations, we obtain explosions in a 1D+ approach employing the Supernova Turbulence In Reduced-dimensionality (STIR) model \citep{couch_simulating_2020}. This method effectively accounts for turbulence and convection, closely guided by 3D simulations. We present the first systematic variation of EOS properties in CCSN simulations with the STIR method (but see \citealp{Boccioli:2021fjh} for a broad EOS exploration with STIR). Since the nucleon effective masses were previously found to have the largest impact on the explosion, we focus especially on the new parameterization of the effective masses. The nucleon effective masses govern the thermal part of the EOS, but as a global parameter of the EDF, they impact both the cold and thermal contributions to the energy and, consequently, the pressure, entropy, and chemical potentials. We investigate the impact of the new effective mass parametrization by comparing to the commonly used Skyrme mean-field effective mass.

This paper is structured as follows. In Section~\ref{sec: method}, we explain how we build the EOS tables and which values we choose for our EOS parameter variations. We also introduce the setup that we employ for the CCSN simulations. In Section~\ref{sec: cold properties}, we present the results of varying cold nuclear matter properties and in Section~\ref{sec: effective mass} the impact of the new effective mass behavior compared to typical mean-field expectations. Finally, we summarize our results in Section~\ref{sec: conclusions}.

\section{New Equation of State Tables}
\label{sec: method}

In this section, we describe the framework used to construct the EOS tables presented in this work, based on the approaches of \cite{huth_new_2021} and \cite{schneider_open-source_2017}. We then summarize the properties of the resulting EOS tables and discuss how they satisfy current nuclear and astrophysical constraints.

\subsection{Framework to build EOS tables}
\label{subsec:eos}

We have incorporated the EDF and effective mass parameterization developed in \cite{huth_new_2021}, hereafter \hws, into the \code\ code, a merged and updated version of the \sro\ and \apr\ codes of \cite{schneider_open-source_2017,schneider_akmal-pandharipande-ravenhall_2019}. In this work, we also use this new framework for the first time in CCSN simulations. 
Our new EOS tables are in agreement with ab initio constraints as described in \cite{huth_new_2021} as well as the updated NICER constraint from \cite{Rutherford:2024srk}. 

The \sro\ and \apr\ codes were developed to compute finite-temperature EOSs of dense matter for astrophysical simulations \citep{schneider_open-source_2017, schneider_akmal-pandharipande-ravenhall_2019} based on the models of \cite{lattimer_generalized_1991} and \cite{Akmal:1998cf}, respectively.
The new \code\ code\footnote{Available after publication.}
includes the \hws\ parameterization of \cite{huth_new_2021} with updates discussed in this work, as well as the capabilities of \sro\ and \apr\ codes. 

All three models assume that nuclear matter is composed of nucleons immersed in an overall charge neutral, degenerate, and non-interacting lepton and photon gas in thermal equilibrium. 
Nuclear matter may phase separate following the liquid-drop model of \cite{lattimer_generalized_1991} and form heavy nuclei immersed in a dilute gas of nucleons and alpha particles, which are treated by an excluded volume approach. 

\begin{figure}[t!]
    \includegraphics[width=0.45\textwidth]{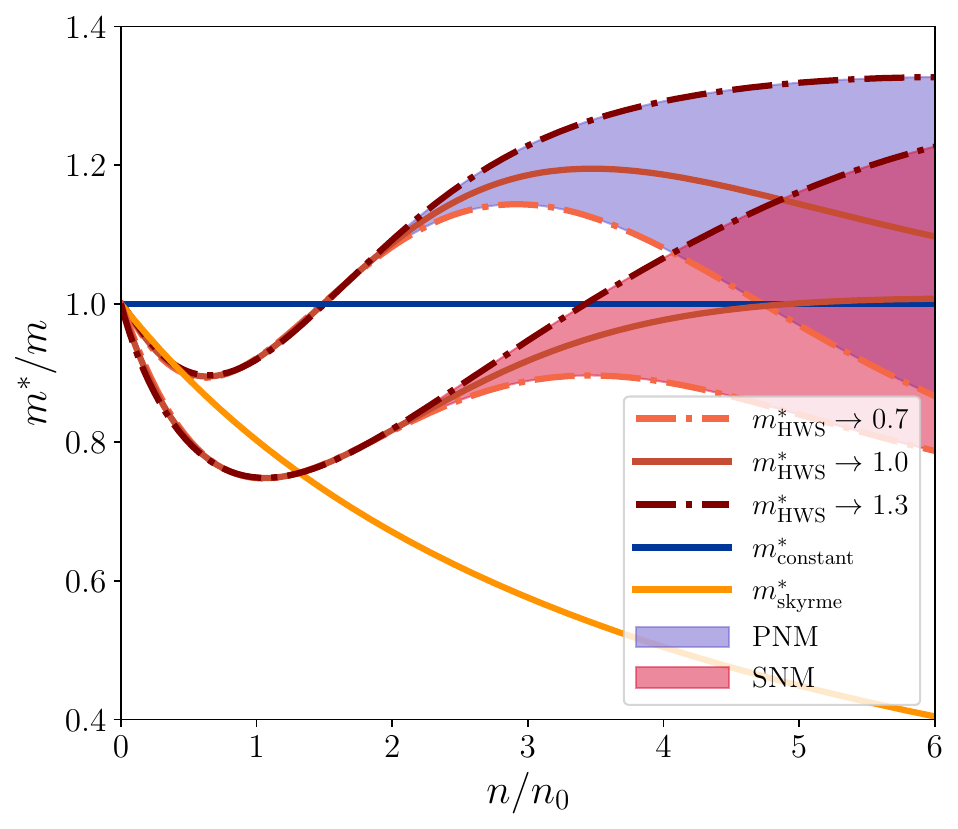}
    \caption{Parametrizations of the effective nucleon mass in PNM and SNM depending on density, compared to a constant $m*_{\rm constant} = m$ and a typical Skyrme-like effective mass. The parametrization $m^*_{\text{HWS}}$ of \cite{huth_new_2021} includes a variation at high density.}
    \label{fig: meff}
\end{figure}

The thermodynamics of baryons ($B$) is described from the energy density of nucleons, which has a kinetic and a potential term: 
\begin{equation}
    \epsilon_B(n,y,T) = \epsilon_{\rkin}(n,y,T) + \epsilon_\rpot(n,y)\,.
\end{equation}
Here, $n=n_n+n_p$ is the baryonic number density, with $n_n$ and $n_p$ the neutron and proton number densities, respectively, while $y=n_p/n$ is the proton fraction of the system and $T$ is the temperature. 
The kinetic energy term has a non-relativistic form 
\begin{equation}
    \epsilon_{\rkin}(n,y,T) = \frac{1}{n}\left(\frac{\hbar^2\tau_n}{2m_n^*} + \frac{\hbar^2\tau_p}{2m_p^*}\right)\,,
\end{equation}
where $\tau_t = \tau_t(n,y,T)$ sets the temperature dependence of the kinetic energy. 
In Skyrme-type EOS models, such as that of \citet{lattimer_generalized_1991}, the effective masses $m_t^*$ are set by
\begin{equation}\label{eq: mstar skyrme}
    \frac{\hbar^2}{2m_t^*} = \frac{\hbar^2}{2m_t} + \alpha_1 n_t + \alpha_2 n_{-t}\,,
\end{equation} 
with $m_t$ the bare nucleon mass, and $\alpha_1$ and $\alpha_2$ parameters of the model where $-t=p$ if $t=n$ and vice versa. 
This results in a simple monotonic (usually decreasing) behavior of the effective mass with increasing density for a fixed composition. 
The \hws\ functional of \citet{huth_new_2021} allows for general effective mass behaviors through
\begin{align}\label{eq: mstar hws}
    \frac{m_t^*}{m_t} &= 1 + \sum_{i=0}^2\frac{\alpha_i n_t^{1+i/3} + \beta_i n_{-t}^{1+i/3}}{1+e^{5n}} \nonumber \\
    &\quad + \left(\epsilon_t\frac{n_t}{n} + \epsilon_{-t}\frac{n_{-t}}{n} - 1\right)\frac{1-e^{-10n}}{1+e^{-5(n-n_{\text{off}})}}\,,
\end{align}
for baryonic densities $n$ in units of \unit{\fm\tothe{-3}}. 
The first term on the right-hand side reflects the bare nucleon mass, while the terms in the summation control the effective mass behavior at low to intermediate densities ($n\lesssim6n_0$ with saturation density $n_0$), with parameters $\alpha_i$ and $\beta_i$ adjusted to reproduce the effective mass behaviors in symmetric nuclear matter (SNM) and pure neutron matter (PNM) obtained by \citet{carbone_ab_2019}. 
Finally, the last term contains the parameters $\epsilon_t$, $\epsilon_{-t}$, and $n_{\text{off}}=\qty{0.7}{\fm\tothe{-3}}$, which determine the high-density behavior of the effective mass. 
We follow \citet{huth_new_2021} and set $\epsilon_n=\epsilon_p$ so that protons and neutrons have the same effective mass in the high-density limit (see Fig.~\ref{fig: meff}). 

We note that the effective mass displays a more complex behavior in the \hws\ model than in \sro\ or \apr\ models, the latter being similar to the Skyrme functional form with the $\alpha_i n_t$ terms multiplied by a decaying exponential term of the form $e^{-p_4n}$, with $p_4>0$ a parameter of the model. 
Figure~\ref{fig: meff} shows a comparison for an effective mass using the \sro\ parameterization, $m^*_{\text{Skyrme}}$, and the limits determined by \citet{huth_new_2021} for the \hws\ model. Despite its more complicated functional form, implementation of the \hws\ model in either the \sro\ or \apr\ codes required only updates to the modules that compute quantities involving the effective masses and nucleonic potential energy density, as well as their derivatives with respect to density $n$ and composition $y$. 

In all three models (\sro, \apr, and \hws), the potential energy is independent of temperature. 
For the \hws\ functional, the potential term has the form
\begin{align}\label{eq:epot}
    \epsilon_\rpot(n,y) = - x\Delta + \sum_{i=1}^4\bigg[& \frac{a_i}{d_a + n^{(\delta_i-2)/3}} \nonumber \\
    & + \frac{b_iy(1-y)}{d_b + n^{(\delta_i-2)/3}}\bigg]n^{\delta_i/3} \,.
\end{align}
Above, $\Delta=m_n-m_p$ is the neutron-proton mass difference, as we measure the energy of the system with respect to a gas of non-interacting neutrons. 
The coefficients $a_i$, $b_i$, $d_a$, $d_b$, and $\delta_i$ are parameters of the model, which can be adjusted to reproduce the properties of nuclear matter. 
In this work, we follow \citet{huth_new_2021} and set $\delta_n=(3,6,9,12)$ with $d_a = d_b = 0.4$ and define a baseline EOS where the coefficients are adjusted to obtain the expected values of some properties of nuclear matter, see Table~\ref{tab:eos} and the discussion in Sec~\ref{subsec: variations}. 
The \sro\ Skyrme form for the nuclear potential is similar to the one in Eq.~\eqref{eq:epot}, setting the denominators in the summation terms of Eq.~\eqref{eq:epot} to one. Note that in the \hws\ model the denominators ensure that the EOS is consistent with the higher-density astrophysics constraints.
The \apr\ nuclear potential includes terms that are polynomial in density, as in the Skyrme case, and exponentially decreasing terms. \apr\ has two different parametrizations, one favored at lower densities and another favored at higher densities, see Section II.A in \citet{schneider_akmal-pandharipande-ravenhall_2019}. 

Besides the updates discussed above, in the region where nuclear matter can cluster into heavy nuclei, \code\ also has a new method to compute the critical temperature $T_\text{crit}(y_i)$ above which heavy nuclei with proton fraction $y_i$ dissolve. 
In \sro\ and \apr\ the stability condition of asymmetric matter was computed using the same procedure as that of symmetric nuclear matter. 
However, this has an effect of lowering the temperature at which nuclear matter becomes unstable. 
Meanwhile, in the improved \code\ code, we follow the methods of \cite{muller_phase_1995} and \cite{avancini_spinodal_2006} to determine the critical temperature $T_\text{crit}(y_i)$ for which asymmetric nuclear matter becomes unstable. 
The critical temperature $T_\text{crit}(y_i)$ is expressed as a fourth-order polynomial in $\delta_i^2=(1-2y_i)^2$. 
In our calculations of the critical temperature, we do not consider the effects of an electron background, which could decrease the region where nuclear matter phase separates and forms nuclei, see discussion by \cite{avancini_spinodal_2006}. 

\begin{table*}[t!]
    \centering
    \begin{tabular}{cccccccccccc}
    $n_0$ [$\si{\per\cubic\femto\metre}$] & $B/A$ [\si{\mega\electronvolt}] & $K$ [\si{\mega\electronvolt}] & $E_\text{sym}$ [\si{\mega\electronvolt}] & $L$ [\si{\mega\electronvolt}] & $P_\text{SNM}$ [$\si{\mega\electronvolt\per\cubic\femto\metre}$] & $P_\text{sym}$ [$\si{\mega\electronvolt\per\cubic\femto\metre}$] & $\delta_1$ & $\delta_2$ & $\delta_3$ & $\delta_4$ & $d_n$ \\
    \hline
    0.164  & 15.86 & 215 & 34 & 55 & 600 & 200 & 3 & 6 & 9 & 12 & 0.4
    \end{tabular}
    \caption{Nuclear matter properties and additional parameters defining the EDF of our baseline EOS (see Section~\ref{subsec:eos}). The baseline EOS includes the new effective mass parametrization of \citep{huth_new_2021} with the asymptotic value $m^*/m\rightarrow1.0$ towards high density.}
    \label{tab:eos}
\end{table*}

As in \cite{lattimer_generalized_1991, schneider_open-source_2017, schneider_akmal-pandharipande-ravenhall_2019, schneider_equation_2020}, the mass and charge of heavy nuclei are determined by employing the single nucleus approximation (SNA) in the liquid-drop model framework. 
These calculations require knowledge about the surface tension of nuclear matter, which depends on the contributions of density gradients to the energy density of nuclear matter, see Equation~(24) in \cite{schneider_akmal-pandharipande-ravenhall_2019}. 
Since the \hws\ model does not include such terms, we use the parameterization of \cite{lattimer_generalized_1991} for the surface tension in terms of the temperature $T$ and proton fraction of the SNA heavy nucleus $y_i$:
\begin{equation}
    \sigma(T,y_i) = \sigma_0 \left[1-\left(\frac{T}{T_\text{crit}(y_i)}\right)^2\right]^p\frac{2 \cdot2^\lambda + q}{y_i^{-\lambda} + q + (1-y_i)^{-\lambda}}\,.
\end{equation}
While the critical temperature $T_\text{crit}(y_i)$ is determined for each EDF parameterization used, the other parameters were fixed to those used by \cite{schneider_equation_2019} for the \sro\ EOSs: $\sigma_0 = \qty{1.15}{\MeV\per\fm\squared}$, $p=\num{1.5}$, $\lambda=\num{3}$, and $q=\num{16}$. 
We ran a few tests with EOSs that differed only in the surface tension parameters and observed that, except for the explosion energy which could differ by a few percent depending on the surface tension parameterization, all other observables differed by less than \qty{1}{\percent}. 
However, our tests were not exhaustive, and we do not discard that different parameterizations of the surface tension could result in larger differences across simulations. 

Finally, as in \cite{schneider_open-source_2017, schneider_equation_2019}, our EOSs transition from the SNA framework to a nuclear statistical equilibrium (NSE) framework considering 3335 nuclei between $-4\lesssim\log_{10}(n\,[\unit{\per\fm\cubed}])\lesssim-2$ centered at $\log_{10}(n\,[\unit{\per\fm\cubed}])=-3$, where both EOSs contribute equally. Thus, for $\log_{10}(n\,[\unit{\per\fm\cubed}])\gtrsim-2$ the EOS comes purely from the SNA component of the model and for $\log_{10}(n\, [\unit{\per\fm\cubed}])\lesssim-4$ from NSE.

\subsection{EOS variation}
\label{subsec: variations}

The introduced framework allows for a systematic variation of nuclear matter properties. For this, we choose values for 8 nuclear matter properties and fit the 8 free parameters of the EDF ($a_i$ and $b_i$) to those. In the following, we explain our choices of nuclear matter properties due to a combination of theoretical, experimental, and observational constraints on the nuclear EOS. In Table~\ref{tab:eos}, we present the nuclear matter properties and additional functional parameters $\delta_i$ and $d_n$ of our baseline EOS. This EOS (as most of the other EOSs we provide) is consistent with all employed constraints. We detail these constraints in the following. 

For all EOSs, we match the energy per particle of PNM at $n = \SI{0.05}{\per\femto\metre\cubed}$ to the quantum Monte Carlo result obtained in \cite{gezerlis_low-density_2010}: $E /N (n=\SI{0.05}{\per\femto\metre\cubed}) = \SI{2.1}{MeV}$. In \cite{drischler_chiral_2019}, the saturation density was constrained to $n_0=\SI{0.164(7)}{\per\femto\metre\cubed}$ and the binding energy per particle at saturation to $B/A=\SI{15.86(57)}{\mega\electronvolt}$. Since these nuclear matter properties are rather well constrained, we fix them to these central values. We vary individually the incompressibility $K$, the symmetry energy $E_\text{sym}$, and the slope parameter $L$ within their uncertainties (see \citet{huth_new_2021} for the definition of these parameters). For the incompressibility $K$, we adopt the uncertainty range of \cite{hebeler_improved_2011} of $K=\SI{215(40)}{\mega\electronvolt}$. A similar result of $K = (182–254) \SI{}{\mega\electronvolt}$ was obtained in \cite{drischler_asymmetric_2016} and also heavy-ion-collision experiments suggest a similar range \citep{lefevre_nuclear_2026}. Based on these constraints, we adopt the following values for $K$ as in \cite{huth_new_2021}: $K\in\{175,215,255\}\,$MeV. 

The symmetry energy $E_\text{sym}$ and the slope of the symmetry energy $L$ are known to be correlated and have been constrained in many ways (see, e.g., \citealp{gandolfi_maximum_2012, hebeler_equation_2013, roca-maza_neutron_2015, tews_symmetry_2017, drischler_how_2020, lattimer_constraints_2023, Lim:2023dbk}). In particular, a joint analysis of experimental, theoretical, and observational constraints by \cite{lattimer_constraining_2013} suggests that these parameters are in
the ranges $\SI{29.0}{\mega\electronvolt}<E_\text{sym}<\SI{32.7}{\mega\electronvolt}$ and $\SI{40.5}{\mega\electronvolt}<L<\SI{61.9}{\mega\electronvolt}$. A more recent analysis including neutron skin measurements from PREX-II \citep{adhikari_accurate_2021} and CREX \citep{adhikari_precision_2022} resulted in $E_\text{sym}=\SI{30.8\pm1.5}{\mega\electronvolt}$ and $L=\SI{40\pm8}{\mega\electronvolt}$ \citep{lattimer_constraints_2023}. A Bayesian uncertainty analysis of chiral EFT calculations of infinite nuclear matter produced the following results: $E_\text{sym}=\SI{31.7\pm1.1}{\mega\electronvolt}$ and $L=\SI{59.8\pm4.1}{\mega\electronvolt}$ \citep{drischler_how_2020}. Since predictions for $E_\text{sym}$ and $L$ vary more depending on the method, we adopt slightly wider ranges for the symmetry energy and its slope: $\SI{30}{\mega\electronvolt} \lesssim E_\text{sym} \lesssim \SI{34}{\mega\electronvolt}$ and $\SI{40}{\mega\electronvolt} \lesssim L \lesssim \SI{65}{\mega\electronvolt}$. Although $E_{\mathrm{sym}}$ and $L$ are varied independently, the adopted combinations remain broadly consistent with the correlations suggested by existing theoretical and experimental constraints (see, e.g., \citealp{lattimer_constraints_2023}).

\begin{figure}[t!]
    \includegraphics[width=0.45\textwidth]{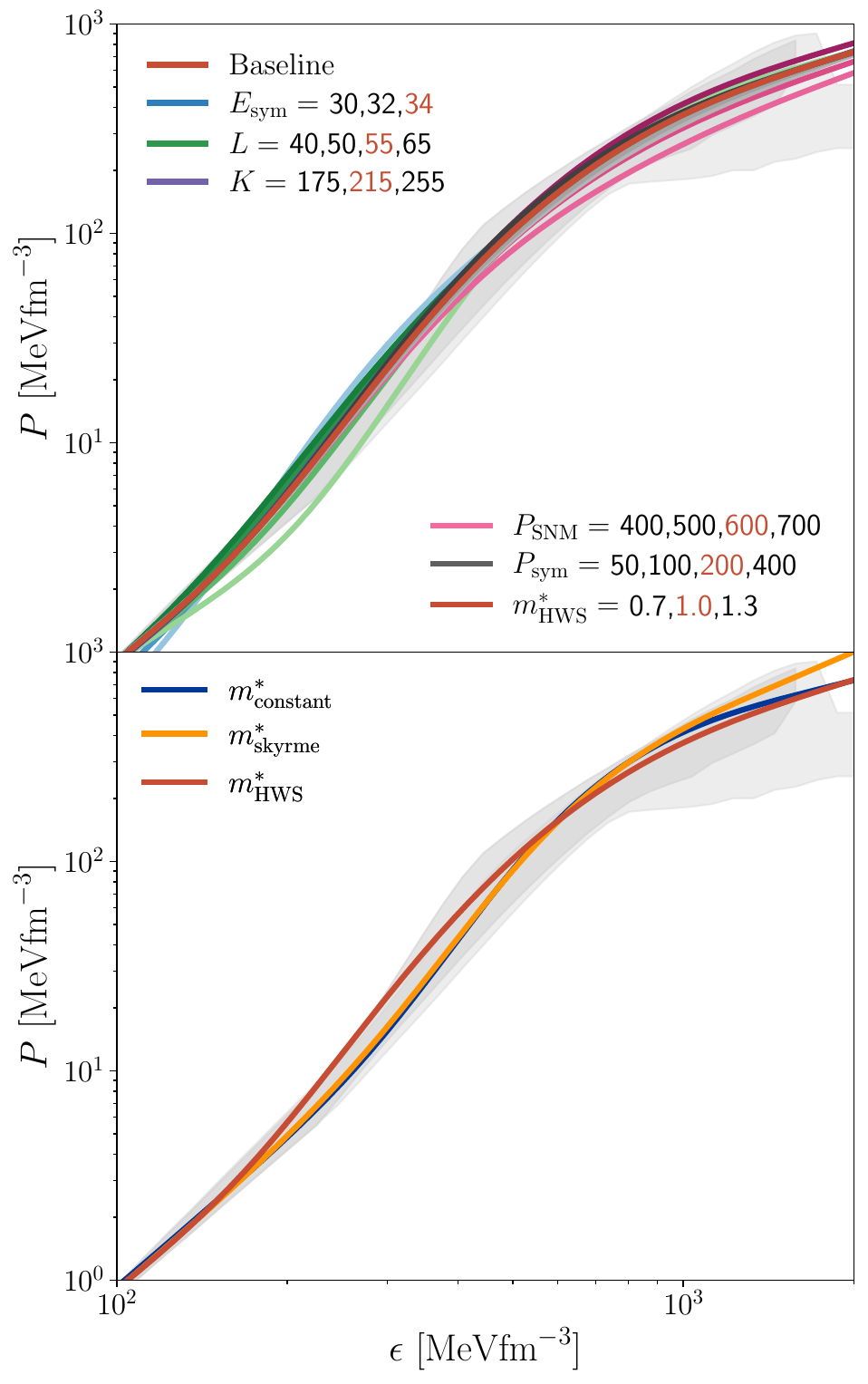}
    \caption{Pressure as a function of energy density of cold, beta-equilibrated neutron star matter for our EOS parameter variations. The lower panel shows the effective mass variation. The two gray bands show the 95\% credible regions of \cite{Rutherford:2024srk} (for the piece-wise polytrope and speed of sound high-density extensions). Our baseline EOS has the values given in Table~\ref{tab:eos} and is marked in red. For each variation, the values given in the legend increase from lighter to darker color.}
    \label{fig: pe}
\end{figure}

\begin{figure}[t!]
    \includegraphics[width=0.45\textwidth]{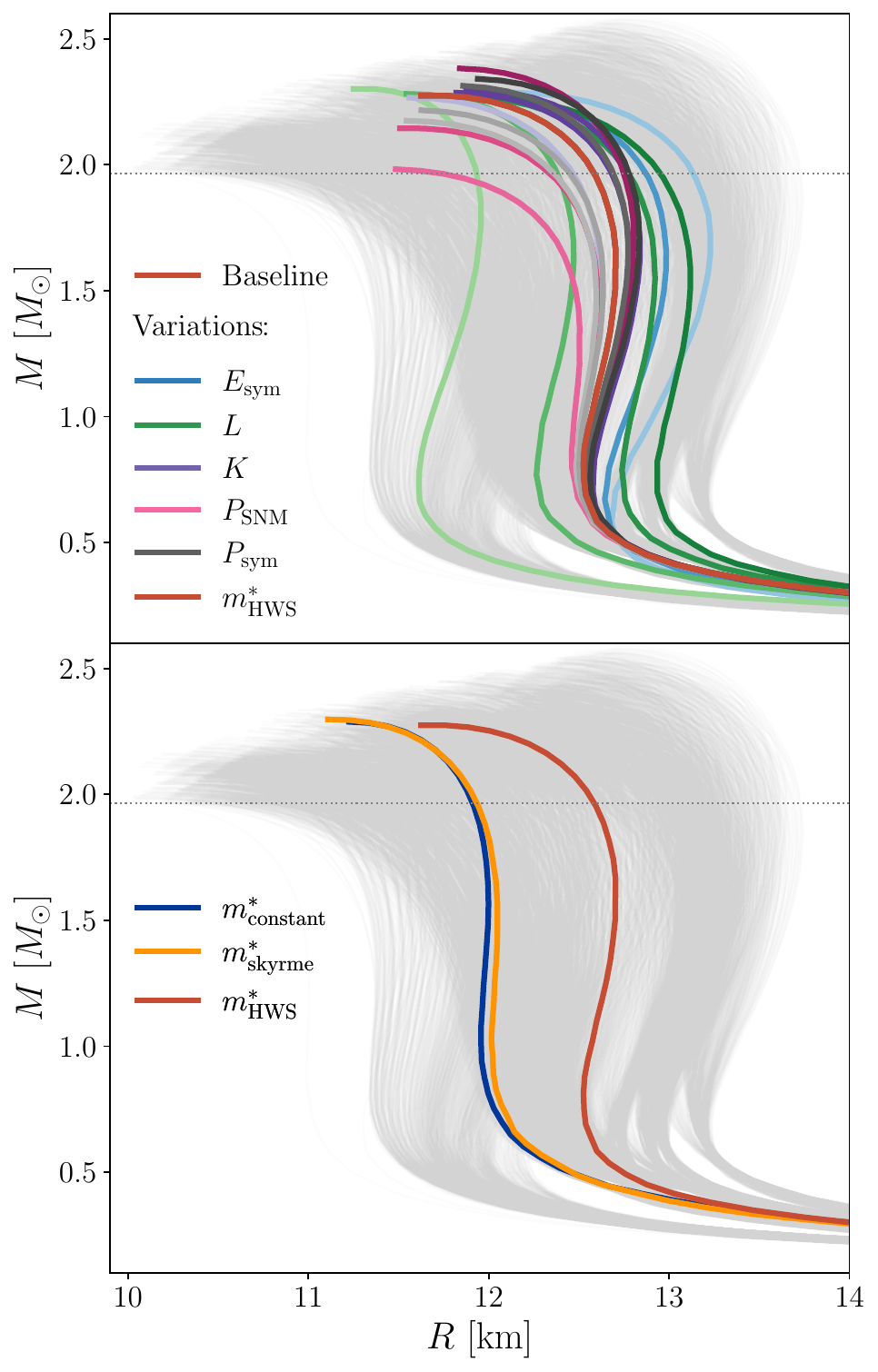}
    \caption{Mass-radius relation of cold, beta-equilibrated neutron stars for our EOS parameter variations. We show all EOSs found to be consistent with all constraints in \cite{huth_new_2021} in gray for comparison and the maximum mass limit $M_\text{NS, max} > 1.9645 M_\odot$ \citep{huth_new_2021} (dotted). The lower panel shows the effective mass variation. Our baseline EOS has the values given in Table~\ref{tab:eos}. For each variation, the values increase from lighter to darker color.}
    \label{fig: mr}
\end{figure}

Additionally, we vary the pressure of PNM and SNM at high density ($n=\SI{1.28}{\per\femto\metre\cubed}\approx 8n_0$). For all EOSs, we employ the astrophysics constraints from heavy pulsar masses, LIGO/Virgo, and NICER from the EOS inference of \citet{Rutherford:2024srk}. In addition, we restrict the pressure to be within 10\% of the theoretical SNM results from functional renormalization group methods by \cite{leonhardt_symmetric_2020}. These constraints allow for a variation of the high-density pressure in the range $P_\text{SNM} = 400-\SI{700}{\mega\electronvolt\per\femto\metre\cubed}$ and $P_\text{sym} = P_\text{PNM} - P_\text{SNM} = 50-\SI{400}{\mega\electronvolt\per\femto\metre\cubed}$. For the effective masses, we employ the three scenarios suggested in \cite{huth_new_2021}, varying the high-density limit to $m^*/m(n\rightarrow\infty) \in \{0.7, 1.0, 1.3\}$.

In Figs.~\ref{fig: pe} and~\ref{fig: mr}, we show the pressure as function of energy density and mass-radius relations of our set of EOS tables for cold neutron star matter in beta equilibrium. Almost all of the parameter combinations are within the pressure band of \cite{Rutherford:2024srk}, except for the EOS with $E_\text{sym} = 30$\,MeV and $\SI{32}{\mega\electronvolt}$, as well as $L = \SI{65}{\mega\electronvolt}$. The one with $E_\text{sym} = \SI{30}{\mega\electronvolt}$ additionally violates the unitary gas constraint \citep{tews_symmetry_2017}. Since these deviations are small, we still keep the EOS. All other EOSs are consistent with all discussed constraints. Additionally, they all fulfill the constraints of causality, and a maximum neutron star mass of at least $M_\text{NS, max} > 1.9645 M_\odot$ (averaged lower bound of \citealp{antoniadis_massive_2013, cromartie_relativistic_2020}) as in \cite{huth_new_2021}.

In all variations, we keep the same baseline properties presented in Table~\ref{tab:eos} and vary only one parameter at the time. We vary $K$, $E_\text{sym}$, $L$, $P_\text{SNM}$, $P_\text{sym}$, and $m^*/m(n\rightarrow \infty)$ within the ranges explained above. This systematic variation allows us to study the effects of individual nuclear matter properties and propagate their uncertainties to the CCSN dynamics.

Furthermore, we investigate the impact of the new effective mass parameterization of \cite{huth_new_2021} with an additional variation, see lower panels in Figs.~\ref{fig: pe} and~\ref{fig: mr}. We produce two additional EOSs with the baseline properties (see Table~\ref{tab:eos}), but with different parameterizations of the effective mass. For the first EOS, we assume that the effective mass is constant and equals the bare nucleon mass: $m^*_\text{constant}=m$. Second, we compare this to the commonly used Skyrme parameterization, where the effective mass is monotonically decreasing with increasing density and $m^*_\text{skyrme}/m=0.75$ at saturation density. We compare these two additional EOSs to our baseline EOS with the new effective mass parameterization: $m^*_\text{HWS}$. This is exactly the same baseline EOS as in the previous variations. Note that changing the parameterization of the effective mass changes the overall behavior of the EDF. To make the models better comparable, we fix, only for these two additional EOSs, the pressure in a relevant density region at $4n_0$ to the value of our baseline EOS instead of fixing the pressure at higher density ($P(\SI{1.28}{\per\femto\meter\cubed})$). The density dependence of the effective mass influences the cold as well as the thermal contribution to the energy and pressure, and we will discuss the impact of both in Section~\ref{sec: effective mass}.

\subsection{Simulation setup}
\label{subsec: setup}

We perform spherically symmetric CCSN simulations utilizing the \textsc{Flash} framework modified for CCSNe in \cite{couch_dependence_2013}, \cite{couch_high-resolution_2014}, and \cite{oconnor_two-dimensional_2018}, including a two-moment neutrino transport that distinguishes between electron neutrinos $\nu_e$, electron anti-neutrinos $\bar{\nu}_e$, and the heavy-flavor (anti-)neutrinos combined $\nu_x$. For the neutrino-matter interactions, we produce tables of emissivities and opacities using the \textsc{NuLib} library \citep{oconnor_open-source_2015} for each EOS table. Note that we have not yet included the nucleon effective masses and self-energies in the calculation of the neutrino rates and plan to do so in the future. We use the default reactions, with one exception. Bremsstrahlung is by default only included for heavy-flavor neutrinos as it is expected to be sub-dominant for electron-type neutrinos. However, we find that bremsstrahlung provides a non-negligible contribution to the absorption opacities and emissivities of electron-type neutrinos and therefore include this reaction for all neutrinos in our tables (see also \citealp{mori_developing_2021}). For the spherically symmetric simulations, we adopt a grid covering $\SI{5e9}{\centi\metre}$ in radius with 11 refinement levels, resulting in minimum grid zones of $\sim\SI{200}{\metre}$. 

To trigger explosions in spherically symmetric simulations, we use the STIR model, developed by \cite{couch_simulating_2020}. In this approach, the strength of convection and turbulence are estimated with a modified mixing length theory and effectively included in the hydrodynamic equations. For all free parameters of the STIR model, we use the suggested values. In particular, for the mixing length parameter $\alpha_\Lambda$, we use their preferred value $\alpha_\Lambda = 1.25$. This value was chosen to match data of a 3D simulation as well as the observed PNS mass distribution.

We show results for two progenitors, one with low compactness ($s20$, $\xi_{2.5} = 0.12$, from \cite{rauscher_nucleosynthesis_2002}, where the compactness $\xi_M$ is defined in \cite{oconnor_black_2011}, and we use our baseline EOS to calculate the compactness at bounce) and a progenitor with high compactness ($s24$, $\xi_{2.5} = 0.44$, from \citealt{woosley_evolution_2002}). We find similar results with both progenitors, but the EOS has more impact on the explosion dynamics in the high-compactness model. The stellar model with low compactness explodes fast (roughly at the same time with all EOSs) and therefore, the center does not reach densities beyond $\SI{7e14}{\gram\per\cubic\centi\metre}$, and the impact of our variations is smaller. In the high-compactness model $s24$, we see significant impact on the PNS properties, explosion timing, and neutrino emission, especially for the variation of the effective mass behavior (see Section~\ref{sec: effective mass}). This model shows a delayed explosion after about $\SI{800}{\milli\second}$ with our baseline EOS and differences in the PNS evolution have more time to affect the explosion. Especially differences between the EOSs at higher densities $n_c>2n_0$ can only show up in this model, where these densities are reached before explosion.

\section{Variation of Nuclear Matter Properties}
\label{sec: cold properties}

\begin{figure}[t!]
    \includegraphics[width=0.45\textwidth]{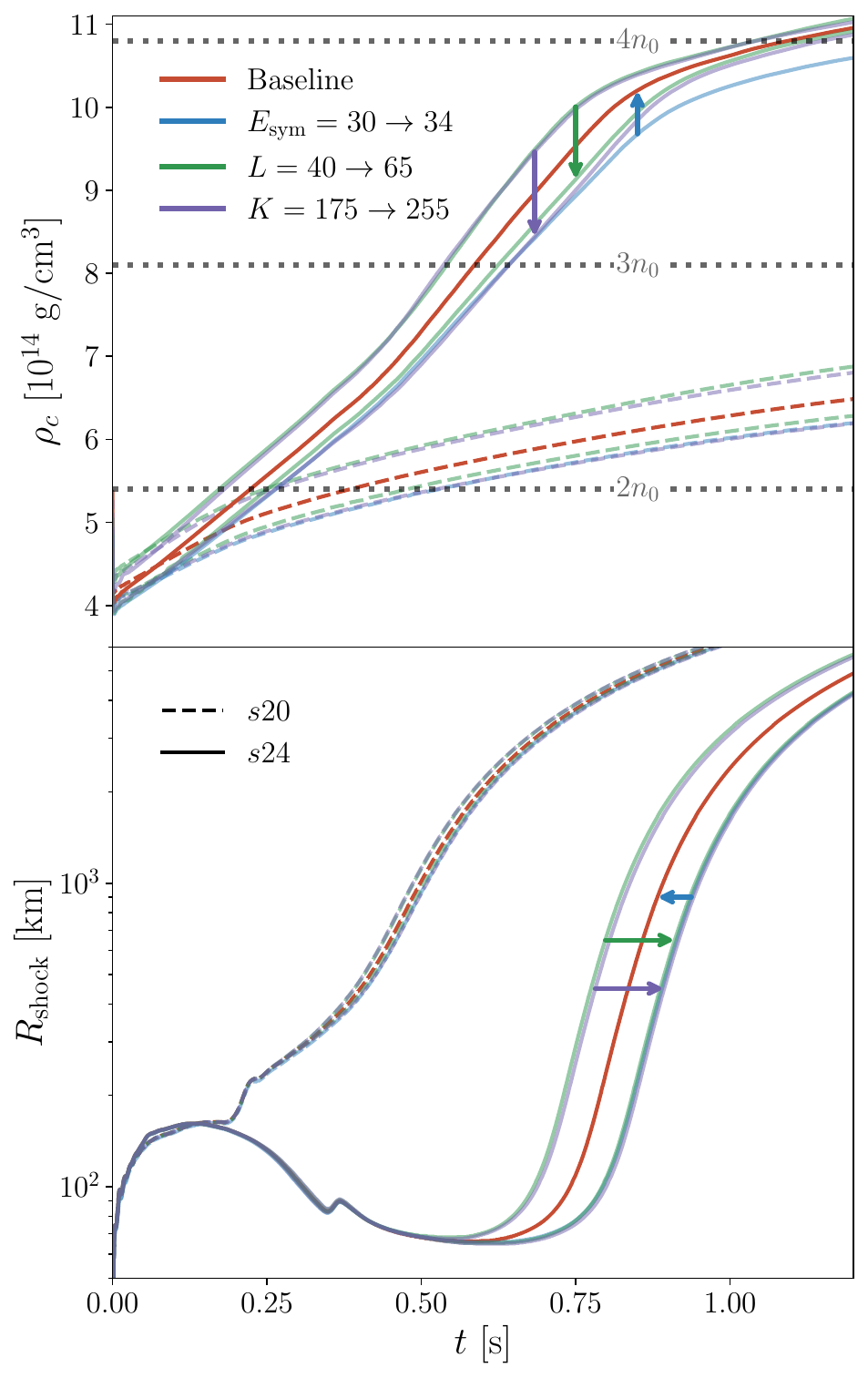}
    \caption{Central density as a function of time (upper panel) and evolution of the shock radius (lower panel) for simulations of the stellar models $s20$ (\citealp{rauscher_nucleosynthesis_2002}, dashed lines) and $s24$ (\citealp{woosley_evolution_2002}, solid lines). The $E_\mathrm{sym}$, $L$, and $K$ variations are shown in blue, green, and purple, respectively. For clarity, we show only the curves corresponding to the lowest and highest value of each parameter. For $s24$, colored arrows indicate how the central density and shock radius change as each parameter is varied from its lowest to its highest value (the $E_\mathrm{sym}$ variation is the smallest arrow). The baseline EOS, with $E_\mathrm{sym}=\SI{34}{MeV}$, $L=\SI{55}{MeV}$, and $K=\SI{215}{MeV}$, is shown in red.}
    \label{fig: Rshock_rhoc_ELK}
\end{figure}

\begin{figure}[t!]
    \includegraphics[width=0.45\textwidth]{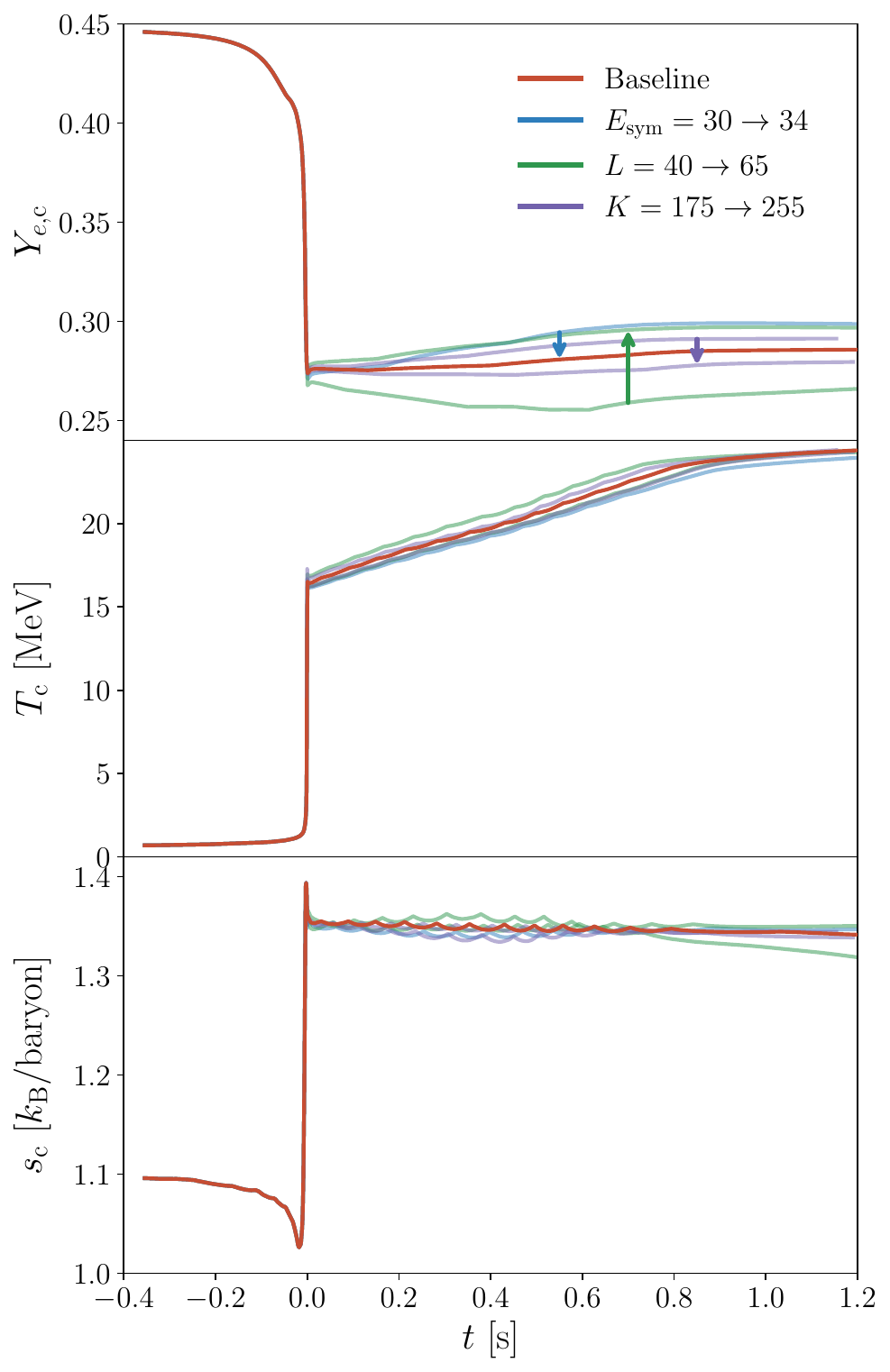}
    \caption{Evolution of electron fraction, temperature, and entropy in the center of the PNS (at $\SI{2}{\kilo\metre}$) as a function of time for simulations of $s24$ (\citealp{woosley_evolution_2002}) with different EOSs varying $E_\text{sym}$, $L$, and $K$. We show the variations of $E_\text{sym}$, $L$, and $K$ in blue, green and purple, respectively. For clarity, we show only the curves corresponding to the lowest and highest value of each parameter. Colored arrows indicate how the central electron fraction changes as each parameter is varied from its lowest to its highest value (the $E_\mathrm{sym}$ variation is the left arrow). The baseline EOS has the values $E_\mathrm{sym}=\SI{34}{MeV}$, $L=\SI{55}{MeV}$, and $K=\SI{215}{MeV}$ and is shown in red.}
    \label{fig: central_ELK}
\end{figure}

\begin{figure}[t!]
    \includegraphics[width=0.45\textwidth]{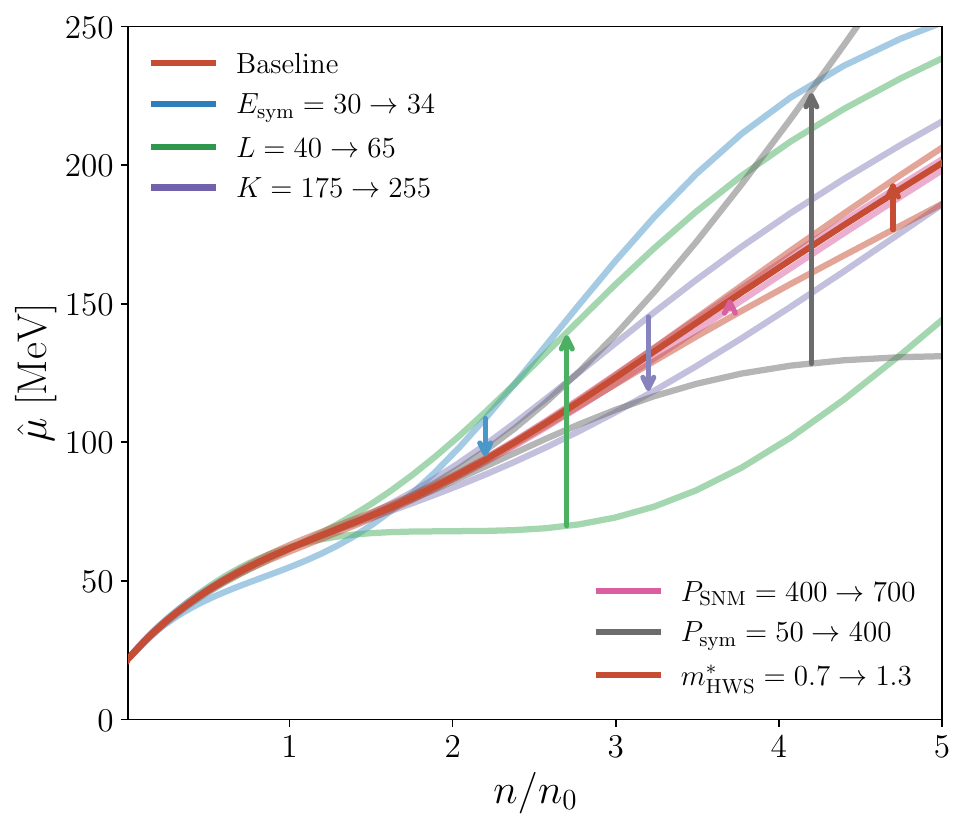}
    \caption{Difference of the neutron and proton chemical potentials as a function of density for different EOSs varying $E_\text{sym}$, $L$, $K$, $P_\text{SNM}$, and $P_\text{sym}$ at $T=\SI{20}{MeV}$ and $Y_e=0.28$. Colored arrows indicate how $\hat{\mu}$ changes as each parameter is varied from its lowest to its highest value. The baseline EOS has the values $E_\mathrm{sym}=\SI{34}{MeV}$, $L=\SI{55}{MeV}$, $K=\SI{215}{MeV}$, $P_\mathrm{SNM}=\SI{600}{\mega\electronvolt\per\femto\metre\cubed}$, $P_\mathrm{sym}=\SI{200}{\mega\electronvolt\per\femto\metre\cubed}$, and $m^*_\mathrm{HWS}=1.0$ and is shown in red.}
    \label{fig: muhat}
\end{figure}

\begin{figure}[t!]
    \includegraphics[width=0.45\textwidth]{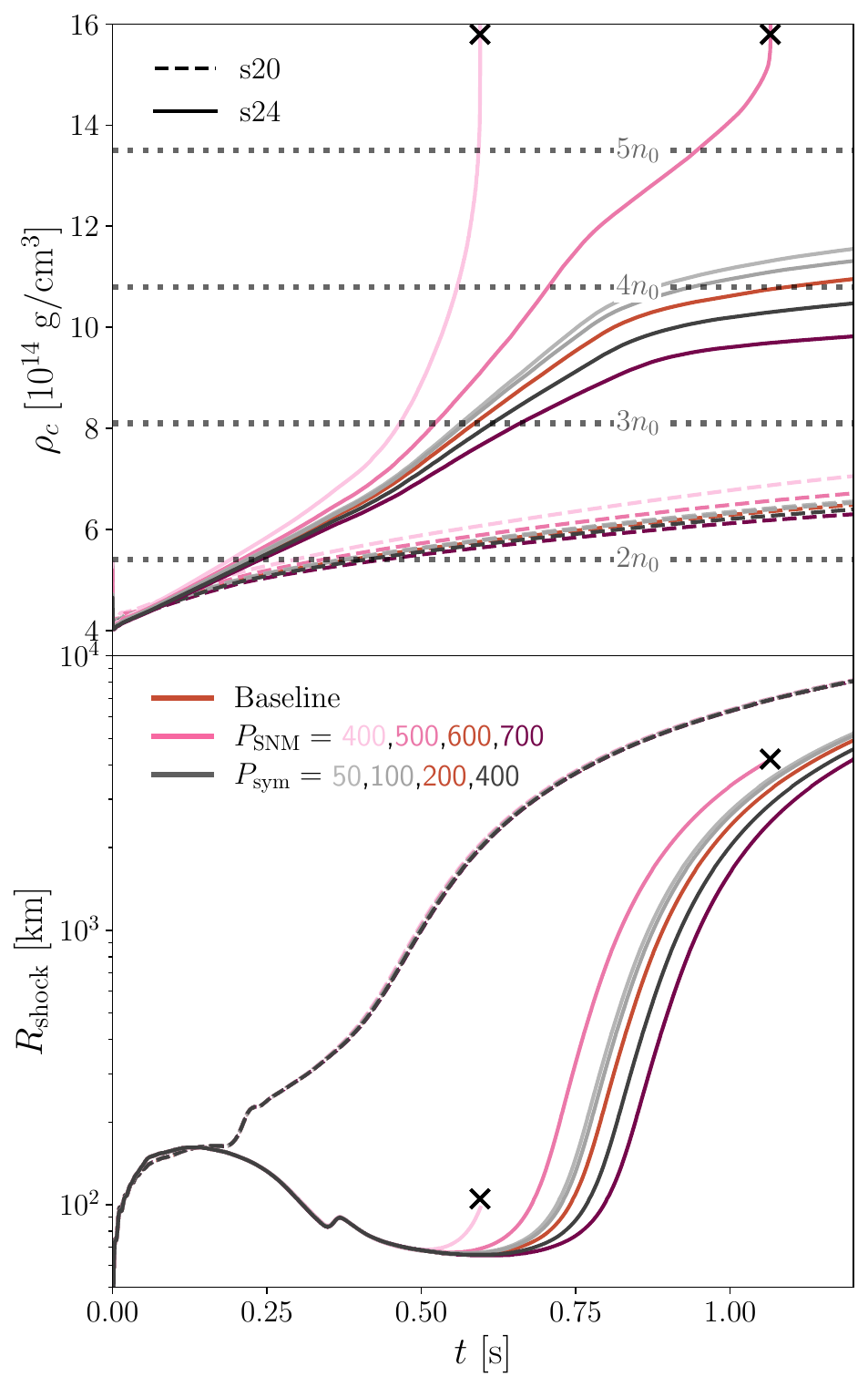}
    \caption{Evolution of central density (upper panel) and shock radius (lower panel) with time  for simulations of the stellar models $s20$ (\citealp{rauscher_nucleosynthesis_2002}, dashed lines) and $s24$ (\citealp{woosley_evolution_2002}, solid lines) with different EOSs varying $P_\text{SNM}$ and $P_\text{sym}$. We show the variation of $P_\text{SNM}$ in pink and $P_\text{sym}$ in gray. For each parameter, the value increases from light to darker color. The baseline EOS has the values $P_\mathrm{SNM}=\SI{600}{\mega\electronvolt\per\femto\metre\cubed}$ and $P_\mathrm{sym}=\SI{200}{\mega\electronvolt\per\femto\metre\cubed}$ and is shown in red. With $P_\text{SNM}=\SI{400}{\mega\electronvolt\per\femto\metre\cubed}$, the star collapses to a black hole after $t_\text{pb}=\SI{0.65}{\second}$ and with $P_\text{SNM}=\SI{500}{\mega\electronvolt\per\femto\metre\cubed}$ after $t_\text{pb}=\SI{1}{\second}$ (marked with a black cross; in the upper panel the corresponding lines continue to even higher densities).}
    \label{fig: Rshock_rhoc_P}
\end{figure}

In this section, we present the results of propagating the uncertainty of cold nuclear matter properties to CCSN explosions. We vary the incompressibility $K$, symmetry energy $E_\text{sym}$, slope parameter $L$, and the pressure at high density as explained in Section~\ref{sec: method}. These parameters all have a rather small impact on the explosion mechanism, consistent with the results of \cite{schneider_equation_2019} and \cite{yasin_equation_2020}. 

The parameter with the smallest effect is the symmetry energy $E_\text{sym}$. As indicated by the blue arrows in Figs.~\ref{fig: Rshock_rhoc_ELK} and~\ref{fig: central_ELK}, there is only a very small difference between the simulations varying $E_\text{sym}$, even for the stellar model with higher compactness $s24$. The $L$ parameter has a larger and opposite effect compared to $E_\text{sym}$ (see green arrows). An increase (decrease) of $L$ ($E_\text{sym}$) translates to higher pressure beyond $n_0$, which leads to a lower central density compared to our baseline EOS. The slightly less compact configuration of the PNS comes with a more shallow mantle, where neutrinos decouple further out at lower temperature. Consequently, the neutrinos are radiated with lower average energy, which reduces the heating rate in the gain layer behind the shock and thus delays the explosion (see lower panel of Fig.~\ref{fig: Rshock_rhoc_ELK}). 

The $L$ parameter additionally influences the rate of deleptonization of the PNS. The value of the electron fraction $Y_e$ depends on the difference between the chemical potentials of neutron and proton $\hat{\mu} = \mu_n-\mu_p$ \citep{bethe_equation_1979} (see Fig.~\ref{fig: muhat}). In the $L$ variation, the post-bounce $Y_e$ varies by less than $0.02$ as $\hat{\mu}$ has similar values for the conditions at bounce. Beyond $1.5n_0$, $\hat{\mu}$ increases more for large $L$, so it takes more energy for the system to increase the isospin asymmetry as the density increases and thus the electron fraction decreases less. Consequently, $Y_e$ differs more for the larger $\hat{\mu}$ because a larger $L$ hinders deleptonization (see green arrow in Fig.~\ref{fig: central_ELK}). The central temperature and entropy are barely affected by these variations. The effect of the symmetry energy and $L$ parameter on the central conditions in the PNS are consistent with what has been obtained in \cite{fischer_symmetry_2014} and \cite{yasin_equation_2020}. The small impact on the shock evolution is in agreement with \cite{schneider_equation_2019} and \cite{yasin_equation_2020}. It is expected, however, that these parameters have a more significant impact in long-term simulations, when the PNS enters in the cooling phase. The deleptonization and convection driven by the lepton fraction gradient depend specifically on the $L$ parameter \citep{roberts_protoneutron_2012}.

The impact of the incompressibility $K$ is of a similar order as that of the $L$ parameter (see purple arrows in Figs.~\ref{fig: Rshock_rhoc_ELK} and~\ref{fig: central_ELK}). A higher incompressibility translates to a higher energy and pressure for SNM beyond $n_0$. Correspondingly, the PNS is more resistant towards compression and stays with a lower central density throughout collapse and the first two seconds of explosion (see Fig.~\ref{fig: Rshock_rhoc_ELK}). Similar to the variation of the $L$ parameter, a lower central density results in a later explosion. The explosion time differs by up to $\SI{150}{\milli\second}$ in the stellar model $s24$, but much less in the low-compactness model $s20$ (see lower panel in Fig.~\ref{fig: Rshock_rhoc_ELK}). While not shown, the delayed explosion with higher $K$ is also a little less energetic. The incompressibility affects the rate at which the PNS deleptonizes as well, but the effect is smaller than in the $L$ variation (see purple arrow in Fig.~\ref{fig: central_ELK}). The difference of the neutron and proton chemical potentials $\hat{\mu}$ is reduced with higher $K$ (see Fig.~\ref{fig: muhat}), so that the PNS can deleptonize faster, starting from the same post-bounce composition. 

The result of later explosions for higher $K$ values is consistent with \cite{couch_dependence_2013}, \cite{schneider_equation_2019} and \cite{yasin_equation_2020}. \cite{eggenberger_andersen_equation--state_2021} have investigated the effect of $K$ using the EOS set from \cite{schneider_equation_2019} in 2D simulations, finding no significant differences in the PNS properties and the neutrino emission. With the turbulent motion in multi-dimensional models, stochastic differences between simulations can be of the same order of magnitude as the differences obtained from nuclear matter uncertainties, washing out the trends observed in 1D simulations.

The variation of the high-density pressure has no effect in the low-compactness model $s20$, where the central density does not exceed $3n_0$, but it does play a role in the high-compactness model $s24$ (see Fig.~\ref{fig: Rshock_rhoc_P}). The lower the pressure at high densities, the faster the central density increases with time. The impact on the explosion timing is on a similar order as for the other variations. For the lowest value of our variation $P_\text{SNM}=\SI{400}{\mega\electronvolt\per\femto\metre\cubed}$, the central density rises so fast in the stellar model $s24$ that the PNS collapses to a black hole, before an explosion is launched (marked with a black cross in Fig.~\ref{fig: Rshock_rhoc_P}). Also with $P_\text{SNM}=\SI{500}{\mega\electronvolt\per\femto\metre\cubed}$, the PNS collapses $\SI{1.06}{\second}$ after bounce (also marked with a black cross), but in this case the explosion is launched before black hole formation. This case can be classified as black hole supernova as explained by \citealp{eggenberger_andersen_black_2025} and is particularly interesting, because the collapse is driven purely by the loss of thermal pressure and not by accretion. The time of black hole formation in collapsing models depends sensitively on the pressure at high densities \citep{schneider_equation_2020}. While the variation of $P_\text{SNM}$ does not have an effect on $Y_e$, a higher $P_\text{sym}$, leads to slightly slower deleptonization of the PNS (see fast increase of $\hat{\mu}$ as function of density for $P_\text{sym}=\SI{400}{\mega\electronvolt\per\femto\metre\cubed}$ in Fig.~\ref{fig: muhat}, indicated with the gray arrow).

The uncertainty propagation of the cold nuclear matter properties presented in this section demonstrates that these variations can have some impact on the explosion timing in a stellar model with relatively high compactness and a delayed explosion. However, in the low-compactness model with early explosion, the impact of these parameters is minor. For the purpose of studying the first two seconds of explosion, the cold nuclear matter properties are sufficiently constrained to not have a major impact on the explosion outcome. Since the newly born PNS is hot, initially thermal effects dominate the EOS impact. We, therefore, focus on these in the following section. 

\section{Impact of the Effective Mass Parametrization}
\label{sec: effective mass}

\begin{figure}[t!]
    \includegraphics[width=0.45\textwidth]{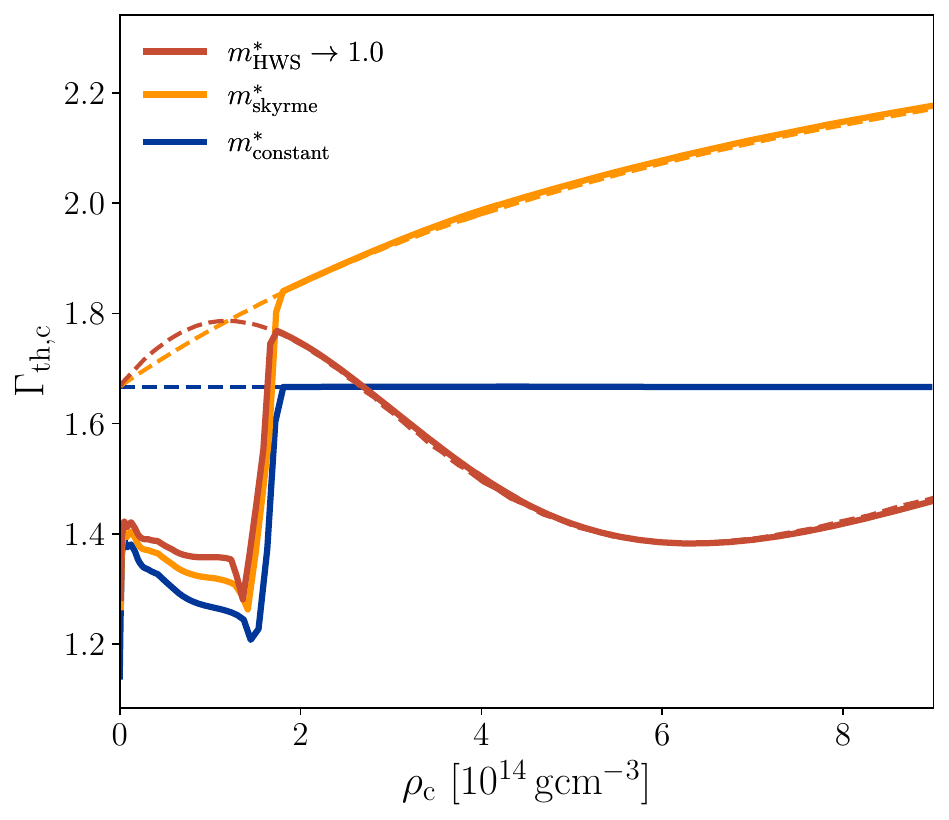}
    \caption{Thermal index as a function of central density for simulations of the stellar model $s24$ (\citealp{woosley_evolution_2002}) with different EOSs varying the parametrization of the effective mass. The thermal index includes only the baryonic contributions (see Eq.~\eqref{eq: gamma_th}, solid lines), and fits exactly to the value of a non-interacting gas of non-relativistic nucleons with effective masses (see Eq.~\eqref{eq: gamma_th approx}, dashed lines) beyond saturation density, where matter is uniform.}
    \label{fig: gammath_mstar}
\end{figure}

A useful parameter to describe the thermal contribution to the baryonic pressure is the thermal index defined as \citep{bauswein_testing_2010,constantinou_thermal_2015}
\begin{equation} \label{eq: gamma_th}
    \Gamma_\text{th} = 1 + \frac{P_\text{th}}{\epsilon_\text{th}} = 1 + \frac{P-P_\text{cold}}{\epsilon-\epsilon_\text{cold}} \,.
\end{equation}
For a Fermi liquid, the thermal index of a non-interacting gas of non-relativistic nucleons at low temperatures can be expressed in terms of the effective mass \citep{constantinou_thermal_2015}: 
\begin{equation} \label{eq: gamma_th approx}
    \Gamma_\text{th} = \frac{5}{3} - \frac{n}{m^*}\frac{\partial m^*}{\partial n} \,,
\end{equation}
with appropriate generalization when the nucleon effective masses differ for neutrons and protons.
We compute the thermal index at the central density of the PNS in our simulations with Eq.~\eqref{eq: gamma_th}, where we read the ``cold'' energy density and pressure from the EOS tables at the minimal temperature of $T=\SI{0.01}{\mega\electronvolt}$. As in \cite{yasin_equation_2020}, we find excellent agreement with the approximation describing the thermal index in terms of the effective masses (Eq.~\eqref{eq: gamma_th approx}) for post-bounce densities (see Fig.~\ref{fig: gammath_mstar}). An accurate description of the effective mass is therefore important to describe the thermal part of the EOS realistically. The effective mass parametrization of \cite{huth_new_2021} is consistent with the chiral EFT calculations up to $2n_0$ of \citep{carbone_ab_2019}. In contrast to typically employed effective mass parametrizations, these show a non-monotonic density-dependence with a minimum around saturation density (see Fig.~\ref{fig: meff}). We show that this behavior has a significant impact on the early explosion dynamics in CCSNe. In the future, it will be interesting to see the effect in simulations of neutron star mergers as well, where thermal effects of the EOS can also affect the dynamics (see, e.g., \citealp{bauswein_testing_2010, hammond_thermal_2021, raithel_realistic_2021, fields_thermal_2023, jacobi_effects_2023}).

\begin{figure}[t!]
    \includegraphics[width=0.45\textwidth]{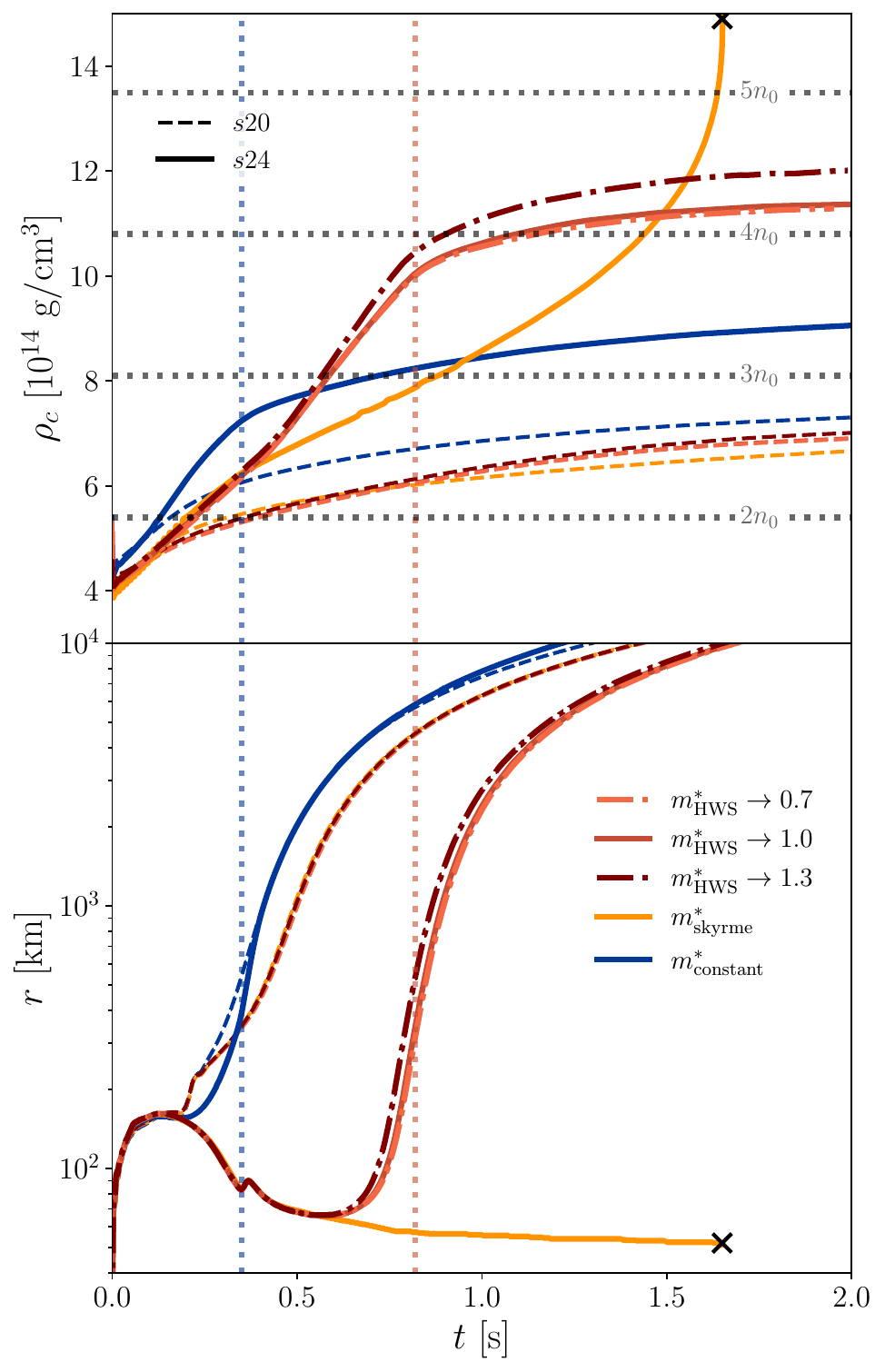}
    \caption{Evolution of central density (upper panel) and shock radius (lower panel) with time for simulations of the stellar models $s20$ (\citealp{rauscher_nucleosynthesis_2002}, dashed lines) and $s24$ (\citealp{woosley_evolution_2002}, solid lines) with different EOSs varying the effective mass. Dotted vertical lines indicate the time of explosion for different models of $s24$, here defined as the time where the diagnostic explosion energy becomes positive. Black crosses mark the time of black hole formation with $m^*_\text{skyrme}$.}
    \label{fig: Rshock_rhoc_mstar}
\end{figure}

\subsection{High-density limit of the effective mass}

The behavior of the effective nucleon masses at higher densities is uncertain and we propagate the different scenarios suggested in \cite{huth_new_2021} to our CCSN simulations (see Fig.~\ref{fig: meff}). We find that the variation of the high-density limit has a very small impact on the explosion as the three scenarios start to differ at $n > 2n_0$ and only significantly differ from $n > 3.5n_0$. These densities are reached only from $\SI{750}{\milli\second}$ after bounce for the stellar model $s24$ and never in $s20$. Still, in model $s24$, the different thermal pressure slightly shifts the time of explosion. With the effective masses asymptoting towards the higher value of $m^*/m \to 1.3$, the thermal pressure increases less with density, so that the PNS can contract a little more. The slightly higher temperature at the neutrinosphere leads to slightly higher luminosities, which makes the shock expand about $\SI{30}{\milli\second}$ earlier (see Fig.~\ref{fig: Rshock_rhoc_mstar}).

\subsection{Parametrization of the effective masses}

\begin{figure}[t!]
    \includegraphics[width=0.45\textwidth]{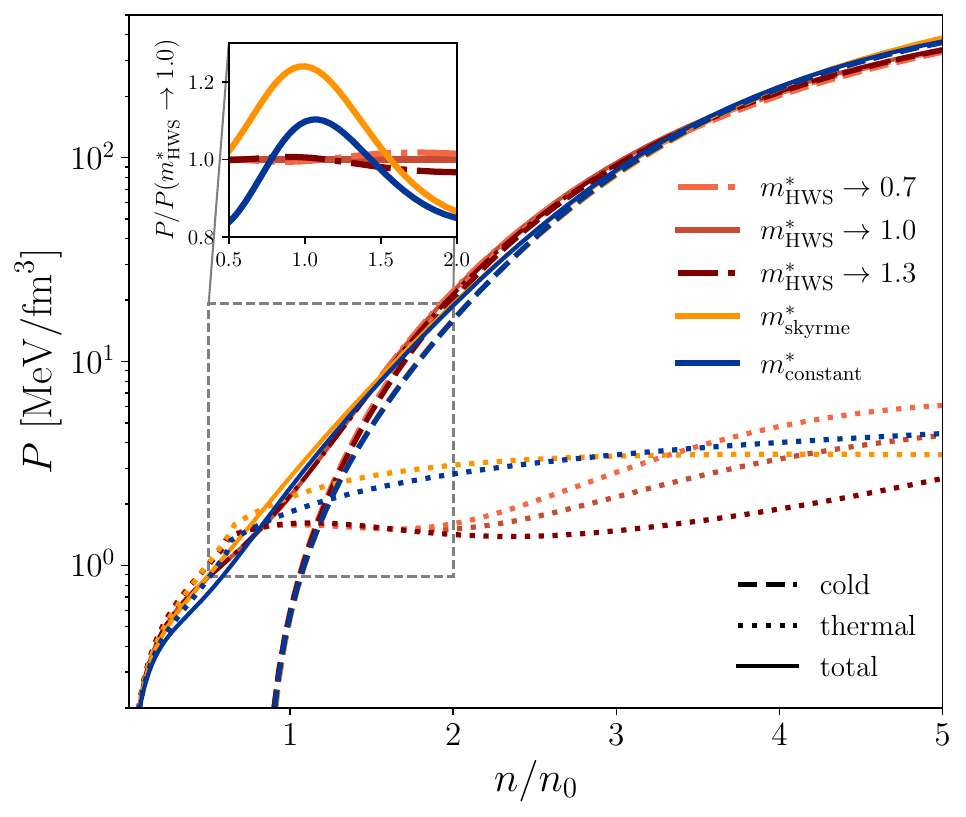}
    \caption{Baryonic pressure of the EOSs with different effective mass parametrizations as a function of number density at $T=\SI{20}{MeV}$ and $Y_e=0.28$. We show the cold (dashed) and thermal (dotted) contributions to indicate where they dominate. At sub-saturation density, the cold pressure is negative, so that the total pressure is lower than the thermal contribution. The pressure is fixed at $4n_0$ to that of the baseline EOS $m^*_{\text{HWS}}\rightarrow1.0$. Note that the pressure does not cross exactly at $4n_0$ as this was fixed for zero temperature for SNM and PNM, not for the conditions displayed here.}
    \label{fig: pressure_mstar}
\end{figure}

The new effective mass parametrization of \cite{huth_new_2021} has a non-monotonic density-dependence, see Eq.~\eqref{eq: mstar skyrme}, featuring a minimum around saturation density consistent with recent chiral EFT calculations \citep{carbone_ab_2019}. We compare this new parametrization to the typically employed Skyrme-type effective mass, see Eq.~\eqref{eq: mstar skyrme}, which monotonically decreases with density and with the case of the effective mass being constant with density and equal to the bare nucleon mass as described in Section~\ref{subsec: variations} (see Fig.~\ref{fig: meff}). The latter is used in the popular EOS of \citet{lattimer_generalized_1991}. Note that the EOSs differ in the cold pressure as well as the thermal contribution. While the cold pressure of the models $m^*_\text{constant}$ and $m^*_\text{skyrme}$ are very similar, the model with the new effective masses has significantly higher cold pressure between $1-4n_0$ and less cold pressure above $4n_0$ (see lower panel in Fig.~\ref{fig: pe}). In Fig.~\ref{fig: pressure_mstar}, we compare the baryonic contributions to the pressure for the different EOSs for typical PNS conditions of $T=\SI{20}{\mega\electronvolt}$ and $Y_e=0.28$. The thermal part dominates the differences between the EOSs up to around saturation density. At higher densities, the cold part of the EOS dominates. We find that these different contributions control the evolution in different regions of the PNS with differences in the cold pressure dominating the central part and thermal effects occurring in the mantle of the PNS. We discuss these effects explicitly in the following.

\begin{figure}[t!]
    \includegraphics[width=0.45\textwidth]{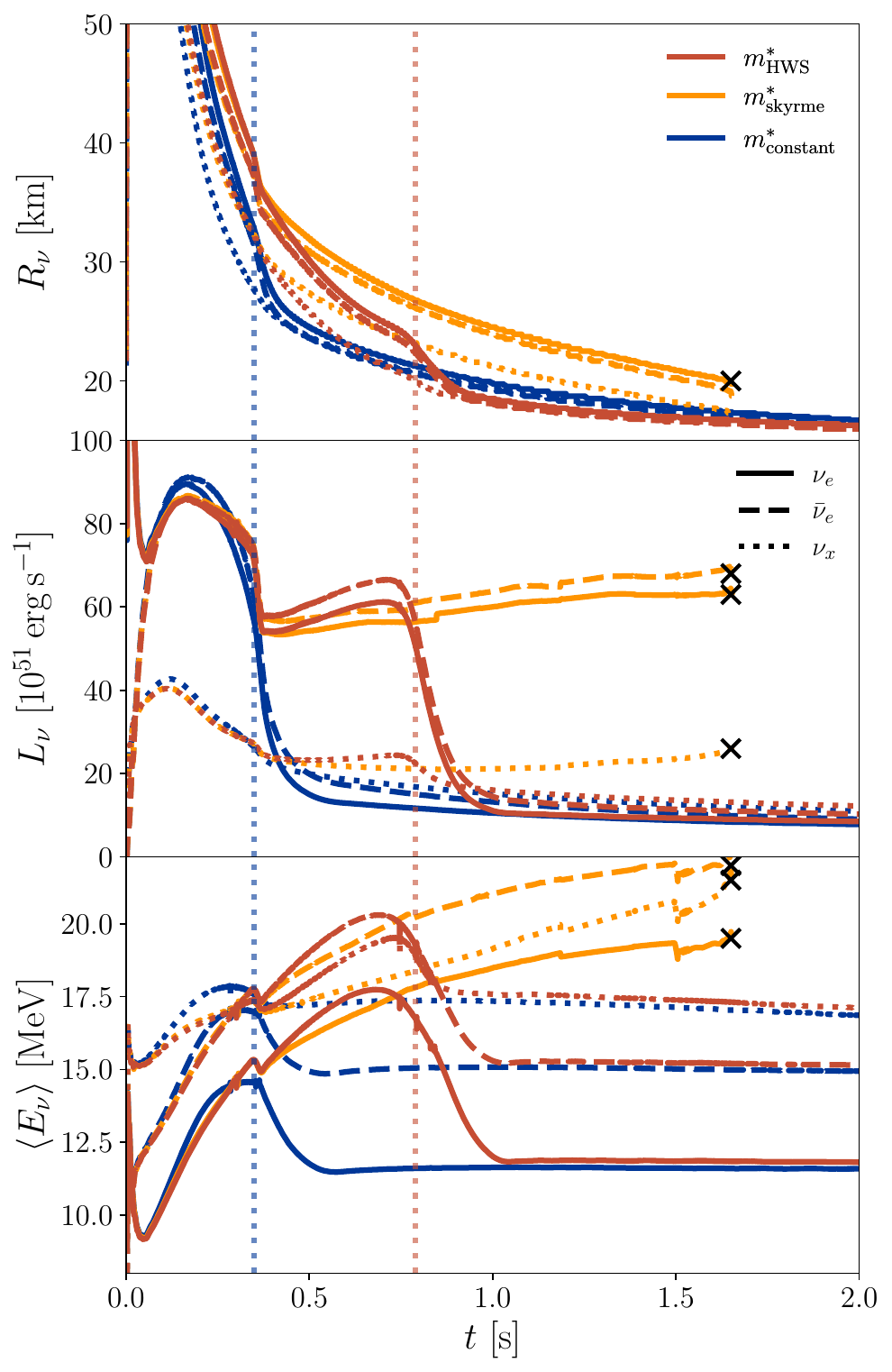}
    \caption{Position of the neutrinospheres (upper panel), neutrino luminosities (middle panel), and average energies (lower panel) as a function of time for simulations of the stellar model $s24$ (\citealp{woosley_evolution_2002}) with different effective mass parameterizations. Solid lines represent electron neutrinos, dashed lines electron anti-neutrinos, and dotted lines heavy flavor neutrinos. Dotted vertical lines indicate the time of explosion for the different models, here defined as the time where the diagnostic explosion energy becomes positive.}
    \label{fig: neutrinos_mstar}
\end{figure}

We start the discussion with model $m^*_\text{constant}$, where the effective nucleon mass is equal to the free nucleon mass for all densities. The PNS of model $m^*_{\text{constant}}$ has a comparably low central pressure and arrives at a compact configuration at bounce ($\rho_c=\SI{4.4e14}{\gram\per\cubic\centi\metre}$, see left panel Fig.~\ref{fig: Rshock_rhoc_mstar}). The contraction behavior of the PNS is governed by the adiabatic index $\Gamma = \partial (\ln P)/\partial(\ln \rho)|_s$. The larger the adiabatic index, the more the system resists against compression. In model $m^*_\text{constant}$, the central density soon exceeds $2n_0$ due to a low adiabatic index between $1-2n_0$ (not shown). The corresponding fast contraction of the PNS moves the neutrinospheres further inside to higher temperatures, leading to higher average energies and luminosities of all neutrinos (see Fig.~\ref{fig: neutrinos_mstar}). The efficient neutrino heating pushes the shock to a larger radius already after $\SI{210}{\milli\second}$ for both stellar models, see Fig.~\ref{fig: Rshock_rhoc_mstar}. Both stellar models investigated have a pronounced drop in density in their pre-collapse profiles, although these density jumps are accreted at different times. In model $s20$, the density jump is accreted by the shock after $\SI{210}{\milli\second}$, while in model $s24$ it is accreted after $\SI{350}{\milli\second}$. The reduced ram pressure with momentarily unchanged neutrino heating allows the shock to move out independent of the EOS (see, e.g., \citealp{boccioli_explosion_2023} for the importance of composition interfaces in stellar progenitors for shock revival). In model $s20$, the explosion is launched in that moment independent of the EOS. However, the shock moves out faster with $m^*_\text{constant}$. We see an even more pronounced EOS dependence in model $s24$. While the shock soon recedes again in the other two effective mass parametrizations due to the reduction of the neutrino luminosities in response to the reduced accretion, the case of $m^*_\text{constant}$ has already gained enough energy after $\SI{350}{\milli\second}$ to launch the explosion (see Fig.~\ref{fig: Rshock_rhoc_mstar}). Since in spherically symmetric models, accretion stops as soon as the shock propagates outward, the central density increases less from that point and the neutrino luminosities and average energies drop (see Fig.~\ref{fig: neutrinos_mstar}).

\begin{figure}[t!]
    \includegraphics[width=0.45\textwidth]{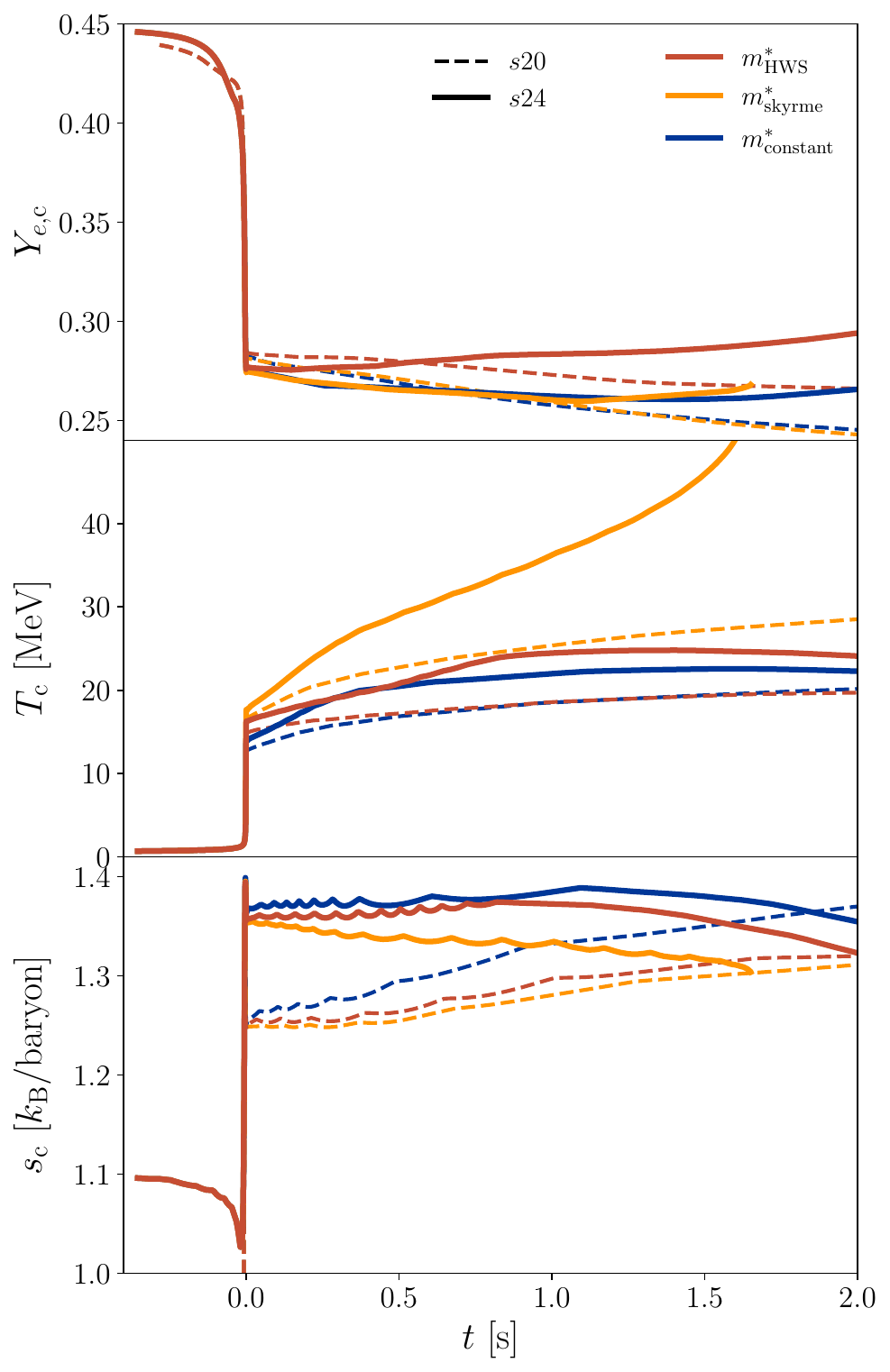}
    \caption{Evolution of electron fraction, temperature, and entropy in the center of the PNS (at $\SI{2}{\kilo\metre}$) for simulations of $s20$ (\citealp{rauscher_nucleosynthesis_2002}, dashed lines) and $s24$ (\citealp{woosley_evolution_2002}, solid lines) with different EOSs varying the effective mass parametrization.}
    \label{fig: central_mstar}
\end{figure}

Next, we discuss $m^*_\text{skyrme}$ with a monotonically decreasing effective mass and correspondingly large thermal index and large thermal pressure (see $\Gamma_\text{th,c}$ in Fig.~\ref{fig: gammath_mstar}). The higher central pressure in this model leads to a lower central density at bounce ($\rho_c=\SI{4.0e14}{\gram\per\cubic\centi\metre}$). The opposite behavior of the central temperature (middle panel in Fig.~\ref{fig: central_mstar}) can be explained assuming Fermi liquid theory in which the entropy scales as $s_c\sim m^* T_c / \rho_c^{2/3} $ \citep{baym_landau_1991} and noticing that the central entropy is only moderately affected by the EOS. As was observed in \cite{schneider_equation_2019} and \cite{yasin_equation_2020}, this configuration leads to a higher central temperature for models with lower effective mass. The same relation explains why the entropy just after bounce is a little lower with $m^*_\text{skyrme}$ and decreases slightly in model $s24$ (see lower panel in Fig.~\ref{fig: central_mstar}), as this model has a decreasing effective mass. Also at higher densities, where the cold pressure dominates, the EOS with $m^*_{\text{skyrme}}$ has the largest cold pressure beyond $4n_0$. Due to the overall higher pressure, the PNS cannot contract as much and stays with a larger radius and consequently the neutrinospheres are located at larger radii (see Fig.~\ref{fig: Rshock_rhoc_mstar}). The correspondingly lower neutrino heating rate is never sufficient to revive the shock and the PNS collapses to a black hole after $\SI{1.65}{\second}$ (marked with black cross in Fig.~\ref{fig: Rshock_rhoc_mstar}). Of course, the EOS dependence of this stellar model is an extreme case and explosions can be obtained with this effective mass parametrization in other models (see results obtained with $s20$ shown with dashed lines). However, this extreme example illustrates well a general behavior. The general trend that EOSs with larger effective masses around $n_0$ (and which thus decrease less with density) yield a hotter neutrinosphere located at smaller radii, aiding shock revival, is in agreement with \cite{schneider_equation_2019} and \cite{yasin_equation_2020} (see also \citealp{rusakov_exploration_2026} for a similar result, although they compare EOSs that differ in more properties than the nucleon effective masses).

Finally, the model with the new effective mass parametrization $m^*_\text{HWS}$ behaves very similar to $m^*_\text{skyrme}$ throughout the evolution of $\qty{2}{\s}$ with $s20$ and also in the first $\SI{300}{\milli\second}$ of evolution with $s24$. However, in model $s24$, their evolution differs and the PNS in $m^*_\text{HWS}$ contracts further to a configuration with an even smaller PNS radius than $m^*_\text{constant}$. Once the density in the PNS center reaches $n_c\approx2n_0$, the adiabatic index of $m^*_{\text{HWS}}$ decreases below the value of $m^*_\text{constant}$ and $m^*_\text{skyrme}$ (not shown) and therefore, the central density keeps increasing more compared to $m^*_\text{skyrme}$ (see Fig.~\ref{fig: Rshock_rhoc_mstar}). Also, in model $s20$, the PNS contracts a bit more with $m^*_\text{HWS}$ compared to $m^*_\text{skyrme}$ once that density is reached. However, in $s24$ this density is reached faster and, since this model did not explode yet, we observe a direct effect of this contraction behavior on the shock evolution. The corresponding recession of the neutrinospheres leads to enhanced neutrino heating, launching a delayed explosion after about $\SI{800}{\milli\second}$. 

\begin{figure}[t!]
    \includegraphics[width=0.45\textwidth]{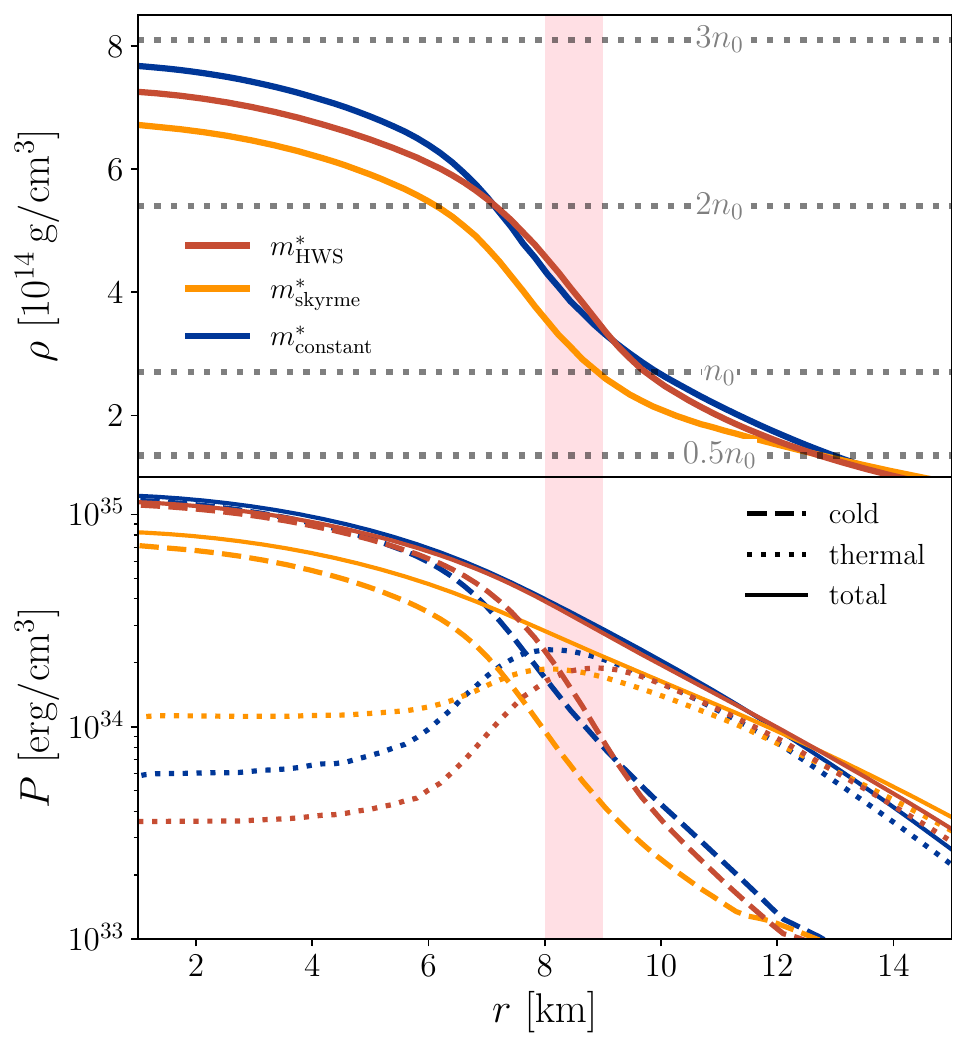}
    \caption{Radial profiles of density (upper panel) and pressure (lower panel) at $\SI{500}{\milli\second}$ post-bounce for simulations of the stellar model $s24$ (\citealp{woosley_evolution_2002}) with different EOSs varying the effective mass parametrization. In the lower panel, we show the ``cold'' pressure of the lowest-temperature slice of the EOS in dashed, the ``thermal'' contribution in dotted, and the total pressure in solid. The region at $8-\SI{9}{\kilo\meter}$, where the thermal pressure starts to dominate with $m^*_\text{HWS}$ is marked in pink. Horizontal dotted lines indicate multiples of the saturation density.}
    \label{fig: profiles_mstar}
\end{figure}

Besides the contraction of the PNS core, the PNS radius and the position of the neutrinospheres are influenced by additional contraction of the PNS mantle. In this region, we can observe a direct effect of the different thermal pressure due to the non-monotonic effective mass. In Fig.~\ref{fig: profiles_mstar}, we show radial profiles of density and pressure $\SI{500}{\milli\second}$ after bounce. As the density increases not only in the center, but throughout the PNS, the PNS mantle around $\SI{9}{\kilo\metre}$ reaches $1-2n_0$. Beyond saturation density, model $m^*_\text{HWS}$ has the lowest thermal pressure (see Fig.~\ref{fig: pressure_mstar}). This is because of the minimum reached in the effective mass around saturation density (see Fig.~\ref{fig: meff}). The increasing effective mass beyond $n_0$ leads to a decrease of the thermal index (see Fig.~\ref{fig: gammath_mstar}) and correspondingly reduced thermal pressure. In the same region, the temperature profile reaches its peak value of up to $\SI{800}{\giga\kelvin}$, enhancing the dominance of thermal effects. In the lower panel of Fig.~\ref{fig: profiles_mstar}, one can see that from around $8-\SI{9}{\kilo\metre}$ outward, the thermal contribution to the pressure is dominating. The lower pressure of model $m^*_\text{HWS}$ in these conditions allows for a stronger contraction of the PNS mantle. The contraction of the PNS mantle in addition to the central contraction has the direct effect of moving the neutrinospheres further inside and thereby enhancing the average energies of radiated neutrinos and neutrino luminosities (see Fig.~\ref{fig: neutrinos_mstar}), which eventually leads to a successful explosion at $\SI{800}{\milli\second}$ after bounce. 

Note that the average energy of electron anti-neutrinos increases to higher values than that of heavy flavor neutrinos for $m^*_{\text{skyrme}}$ and $m^*_{\text{HWS}}$, even though, the neutrinosphere where heavy flavor neutrinos decouple is located at a smaller radius (see Fig.~\ref{fig: neutrinos_mstar}). This can be understood from a non-monotonic behavior of the internal energy and temperature with radius around the neutrinospheres due to neutrino cooling (not shown here, but see \citealp{muller_new_2014} for a discussion). The non-monotonic radial profile of the internal energy makes electron anti-neutrinos decouple on average with higher energy at a larger radius.

The electron fraction in the PNS center is not much affected by the effective mass variation (upper panel in Fig.~\ref{fig: central_mstar}). The difference between the chemical potentials of neutrons and protons $\hat{\mu}$ is a bit larger for $m^*_{\text{HWS}}$ around $2-3.5n_0$. Therefore, the deleptonization is slow and $Y_\text{e}$ stays roughly constant during the simulation time. In both exploding models, the central temperature saturates once the explosion sets in and accretion ceases (similar to \citealp{Boccioli:2021fjh}).

Comparing different EOS frameworks with a simulation setup similar to ours (STIR in GR1D), \cite{Boccioli:2021fjh} find that the EOS which gives the highest central entropy results in the earliest and strongest explosion, a correlation already suggested in \cite{bethe_supernova_1990}. Moreover, \cite{yasin_equation_2020} find that a larger effective mass leads to a higher central entropy together with the model being more prone to explosion. We can observe the same trend in our variation of the effective mass: model $m^*_\text{constant}$, which has the largest effective mass around saturation density, has the highest central entropy just after bounce (see lower panel in Fig.~\ref{fig: central_mstar}) and shows the earliest and most energetic explosion in both stellar models.

As we have seen, the different effective mass parameterizations can result in significantly different evolutions with implications for observables. While we have focused on the neutrino emission, \cite{eggenberger_andersen_equation--state_2021} have investigated the impact of changing the effective mass with the EOS set of \cite{schneider_equation_2019} on the emission of gravitational waves (GWs). They find that the stronger neutrino heating resulting from a higher effective mass, together with more enhanced neutrino-driven instabilities in the gain region, increases the GW amplitude and the energy emitted in GWs. We expect from the above analysis that changing the effective mass parametrization will change the GW signal in a similar or even more pronounced way.

\section{Conclusions}
\label{sec: conclusions}

Recent advances in nuclear experiments, astrophysical observations, and theoretical many-body calculations continue to refine our understanding of dense nuclear matter and its EOS. As these constraints become more narrow, EOS models used in astrophysical simulations must be updated accordingly. In this work, we provide a new set of EOS tables for astrophysical applications that are consistent with recent constraints. Particularly, we have incorporated the new \hws\ EDF and effective mass parametrization of \citealp{huth_new_2021} into the \code\ code and used it to study the effect of systematic variations of individual nuclear matter properties within their constraints on the EOS and astrophysical consequences. Notably, the \hws\ model features a non-monotonic density-dependence of the nucleon effective masses with a minimum around saturation density, which is consistent with recent chiral EFT calculations of \citealp{carbone_ab_2019}.

While our new EOS tables are suitable for different astrophysical environments, we focus here on the impact on CCSN explosions. We propagate the uncertainties of nuclear matter properties to the EOS and the CCSN dynamics. For this, we perform spherically symmetric simulations, following the evolution of the stellar core up to $\qty{2}{\s}$ after core bounce. This is the first systematic propagation of EOS uncertainties to CCSN simulations using the STIR method \citep{couch_simulating_2020} to effectively account for turbulent motion and trigger explosions in spherically symmetric simulations. As mentioned in Section~\ref{subsec: setup}, one should include the nucleon effective masses and self-energies in the calculation of the charged-current neutrino-matter interaction rates to make them consistent with the adopted EOS and we plan to do so in the future. We do not expect that the presented results would change significantly with this update as it modifies the neutrino interactions at densities beyond $\SI{e12}{\gram\per\cubic\centi\meter}$, interior to the neutrino decoupling region (see, e.g., \citealp{martinez-pinedo_charged-current_2012, kotake_impact_2018}). We study the effect of our EOS variations in two stellar models with very different values of core compactness and find a significantly larger impact in the higher-compactness model ($s24$, $\xi_{2.5} = 0.44$, from \cite{woosley_evolution_2002}).

The detailed uncertainty propagation of cold nuclear matter properties shows that these parameters are already reasonably well constrained for the purpose of simulating early CCSN dynamics. The effect of varying the incompressibility, symmetry energy, and $L$ parameter on the PNS evolution is consistent with previous works (see, e.g., \citealp{schneider_equation_2019, yasin_equation_2020}). While there is no significant impact of these variations on the shock evolution in the stellar model $s20$, the more compact model $s24$ shows a delayed explosion, where the variations can lead to a shift of the explosion time of up to $\SI{150}{\milli\second}$. 

Variations of the pressure at high densities propagate in a similar way with a significant impact only in the higher-compactness model. With the exception of the lowest high-density pressure $P_\text{SNM}=\SI{400}{\mega\electronvolt\per\cubic\femto\metre}$, which leads to black hole formation within $\SI{500}{\milli\second}$ and before an explosion is launched, the impact on the explosion is on a similar order as the other variations. 

The main novelty of this study is the impact of the new parameterization of the effective masses. In the early post-bounce evolution, the PNS is hot and thermal effects dominate its evolution. We show that the non-monotonic density dependence of the effective masses has a significant impact on the evolution of the PNS. 

Generally, a higher effective mass allows for faster contraction of the PNS due to a lower thermal pressure and is favorable for explosion (consistent with \citealp{schneider_equation_2019, yasin_equation_2020, boccioli_explosion_2023}). Accordingly, for both stellar models, the model $m^*_\text{constant}$ results in the fastest and most energetic explosion, because for both other parameterizations the effective mass is decreasing up to saturation density. In the stellar model with higher compactness and late explosion, the non-monotonic behavior of the new effective mass parameterization significantly alters the dynamics and outcome. Interestingly, a late explosion is triggered by a faster contraction of the PNS at higher densities ($n_c>2n_0$). Instead, model $m^*_\text{skyrme}$, where the effective mass monotonically decreases with density, fails to explode this stellar model, because the high pressure in the PNS pushes the neutrinospheres too far out to achieve sufficient neutrino heating. We show that the different contraction behavior of the PNS due to the new effective mass parameterization is due to differences in the cold pressure in the PNS center as well as differences in the thermal pressure in the PNS mantle. 

The present study highlights that the parameterization of the EDF and especially the effective masses in phenomenological EOS models has a big impact on CCSN explosions. The impact is significantly larger than for individual parameter variations within current constraints. It is therefore important to use updated models that are well informed by state-of-the-art nuclear theory and experiment. The new effective mass parameterization of the \hws\ model reproduces the ab initio calculations of the finite-temperature EOS and effective masses from chiral EFT from \cite{carbone_ab_2019} up to $\sim 2n_0$ including the minimum around saturation density. The density dependence of the effective masses alters the contraction of the PNS and consequently the simulation outcome, in extreme cases, between a prompt, delayed, or failed explosion.

In the future, it will be interesting to investigate the impact of our new EOS tables in multi-dimensional CCSN simulations and in other astrophysical events, such as neutron star mergers. Multi-dimensional simulations are required to capture the full complexity of the explosion mechanism, including the interplay between the EOS, hydrodynamic instabilities, convection, and other multi-dimensional effects. Additionally, multi-dimensional simulations allow for further multi-messenger predictions including the gravitational wave signal. While the impact of the nuclear properties on the PNS structure should remain the same, the resulting impact on the explosion dynamics and observables may or may not be different. \cite{eggenberger_andersen_equation--state_2021} found enhanced neutrino heating with a higher effective mass in 2D simulations in agreement with previous 1D studies. However, the additional degrees of freedom and statistical variations between simulations also make it more difficult to isolate the impact of individual nuclear properties. One-dimensional simulations therefore remain a valuable complementary approach, as they provide a controlled and computationally efficient framework for systematically exploring EOS variations and identifying the physical origin of their effects. 

Recent studies, such as \cite{powell_impact_2025} and \cite{rusakov_exploration_2026}, have confirmed that the nuclear EOS still has a significant impact in modern multi-dimensional simulations. Large differences are obtained when fundamentally different EOS frameworks are compared. Our new set of EOS tables enables a complementary approach by allowing systematic sensitivity studies and uncertainty propagation of individual nuclear matter properties within a single theoretical framework. While the resulting variations may be more subtle than those obtained when comparing completely different EOS frameworks, this approach allows us to disentangle the contribution of individual nuclear properties to the astrophysical outcome. 

\begin{acknowledgments}
We acknowledge the extensive work of Sean Couch in the development of \textsc{Flash} for CCSN simulations. This work was supported in part by the Deutsche Forschungsgemeinschaft (DFG, German Research Foundation) -- Project-ID 279384907 - SFB 1245. E.O. is supported by the Swedish Research Council (Project No. 2020-00452). G.N. is supported by the grant JDC2024-055535-I funded by MICIU/AEI/10.13039/501100011033 and by FSE+, and acknowledges support from the grant PID2021-127495NB-I00 funded by MCIN/AEI/10.13039/501100011033 and by the European Union, and from the Astrophysics and High Energy Physics programme of the Generalitat Valenciana ASFAE/2022/026 funded by MCIN and the European Union NextGenerationEU (PRTR-C17.I1). Y.L. was supported by the National Research Foundation
of Korea (NRF) grant funded by the Korea government (MSIT)
(No.~RS-2024-00457037) and by the Global--Learning \& Academic Research
Institution for Master's and PhD students, and Postdocs (LAMP) Program of
the National Research Foundation of Korea (NRF) grant funded by the
Ministry of Education (No.~RS-2024-00442483).

\end{acknowledgments}

\software{\code, \textsc{Flash} \citep{fryxell_flash_2000,couch_high-resolution_2014}, NuLib \citep{oconnor_open-source_2015}, Matplotlib \citep{hunter_matplotlib_2007}, NumPy \citep{harris_array_2020}, pandas \citep{mckinney-proc-scipy-2010, the_pandas_development_team_pandas-devpandas_2026}, yt \citep{turk_yt_2010}, h5py \citep{collette_h5pyh5py_2022}}.

\section*{Data Availability}

All EOS tables will be made publicly available online. Further data underlying this study will be provided upon reasonable request.

\bibliography{EOS_variation,added_refs}
\bibliographystyle{aasjournalv7}

\end{document}